\documentclass[
  obeyspaces,
  journal=pasa,
  manuscript=research-paper, 
  year=202X,
  volume=YY,
]{cup-journal}

\usepackage{multirow}
\usepackage{subcaption}
\usepackage{makecell}
\usepackage{amsmath}
\usepackage[skip=0.5ex]{subcaption}
\usepackage{amssymb,siunitx,booktabs,bm}
\usepackage{hyperref}
\hypersetup{colorlinks,citecolor=blue,linkcolor=blue,urlcolor=blue}
\usepackage[normalem]{ulem}
\usepackage{orcidlink}

\newcommand{\de}{\mathrm{d}}

\title{EMU: Cross-correlating EMU Pilot Survey 1 with Dark Energy Survey to validate the radio galaxy bias and redshift distribution}

\author{Chandra Shekhar Saraf\orcidlink{0000-0002-5149-4042}}
\affiliation{Korea Astronomy and Space Science Institute, 776 Daedeok-daero, Yuseong-gu, Daejeon 34055, Republic of Korea}

\author{David Parkinson\orcidlink{0000-0002-7464-2351}}
\affiliation{Korea Astronomy and Space Science Institute, 776 Daedeok-daero, Yuseong-gu, Daejeon 34055, Republic of Korea}
\email[David Parkinson]{davidparkinson@kasi.re.kr}

\author{Jacobo Asorey\orcidlink{0000-0002-6211-499X}}
\affiliation{Departamento de F\'isica Te\'orica, Centro de Astropart\'iculas y F\'isica de Altas Energ\'ias, Universidad de Zaragoza, 50009 Zaragoza, Spain}

\author{Catherine L. Hale\orcidlink{0000-0001-6279-4772}}
\affiliation{Institute for Astronomy, University of Edinburgh, Royal Observatory Edinburgh, Blackford Hill, Edinburgh, EH9 3HJ, UK}
\alsoaffiliation{Astrophysics, Department of Physics, Denys Wilkinson Building, University of Oxford, Keble Road, Oxford, OX1 3RH, UK}

\author{Benedict Bahr-Kalus\orcidlink{0000-0002-4578-4019}}
\affiliation{INAF -- Istituto Nazionale di Astrofisica, Osservatorio Astrofisico di Torino, Via Osservatorio 20, 10025 Pino Torinese, Italy}
\alsoaffiliation{Dipartimento di Fisica, Universit\`a degli Studi di Torino, Via P.\ Giuria 1, 10125 Torino, Italy}
\alsoaffiliation{INFN -- Istituto Nazionale di Fisica Nucleare, Sezione di Torino, Via P.\ Giuria 1, 10125 Torino, Italy}

\author{Maciej Bilicki\orcidlink{0000-0002-3910-5809}}
\affiliation{Center for Theoretical Physics PAS, al. Lotnikow 32/46, 02-668 Warsaw, Poland}

\author{Stefano Camera\orcidlink{0000-0003-3399-3574}}
\affiliation{Dipartimento di Fisica, Universit\`a degli Studi di Torino, Via P.\ Giuria 1, 10125 Torino, Italy}
\alsoaffiliation{INFN -- Istituto Nazionale di Fisica Nucleare, Sezione di Torino, Via P.\ Giuria 1, 10125 Torino, Italy}
\alsoaffiliation{INAF -- Istituto Nazionale di Astrofisica, Osservatorio Astrofisico di Torino, Via Osservatorio 20, 10025 Pino Torinese, Italy}

\author{Andrew M. Hopkins\orcidlink{0000-0002-6097-2747}}
\affiliation{School of Mathematical and Physical Sciences, 12 Wally’s Walk, Macquarie University, NSW 2109, Australia}

\author{Konstantinos Tanidis\orcidlink{0000-0001-9843-5130}}
\affiliation{Astrophysics, Denys Wilkinson Building, Department of Physics, University of Oxford, Keble Road, OX1 3RH, UK}


\begin{document}

\begin{abstract}
Radio continuum galaxy surveys can provide a relatively fast map of the projected distribution of structure in the Universe, at the cost of lacking information about the radial distribution. We can use these surveys to learn about the growth of structure and the fundamental physics of the Universe, but doing so requires extra information to be provided in the modelling of the redshift distribution, $dN/dz$. In this work, we show how the cross-correlation of the two-dimensional radio continuum map with another galaxy map (in this case a photometric optical extragalactic survey), with a known redshift distribution, can be used to determine the redshift distribution through statistical inference. We use data from the Evolutionary Map of the Universe (EMU) Pilot Survey 1 and cross-correlate it with optical data from the Dark Energy Survey to fit the parameters of our $dN/dz$ model. We show that the recovered distribution has a similar shape to the distribution predicted by the current state-of-the-art simulation, and can fit the angular power spectrum data from cross-correlations very well, validating this model. These results will have significance for future cosmological analyses with large-scale radio continuum surveys such as the full EMU, or with the SKAO.
\end{abstract}

\section{Introduction}\label{sec:introduction}

The large-scale structure (LSS) experiments over the last three decades have progressed with a common goal: to understand the formation and evolution of the cosmic web, and to constrain the cosmological parameters that govern it. LSS surveys have played a crucial role in addressing these fundamental questions -- from the dynamics of the inflationary epoch to the nature of the accelerated expansion of the Universe. Different probes of the LSS, such as the weak gravitational lensing (measured either from cosmic microwave background (CMB) or shapes of background galaxies, \citealt{2001PhR...340..291B}), angular clustering of galaxies \citep{2008Natur.451..541G, 2011MNRAS.415.2876B, 2013MNRAS.436.3089B, 2013A&A...557A..54D, 2015MNRAS.449..848H, 2016PASJ...68...38O, 2017A&A...604A..33P, 2022ApJ...928...38T, 2022MNRAS.517.3407M, 2024MNRAS.527.6540H}, and their combinations \citep{2015MNRAS.449.4326P, 2018ApJ...862...81B, 2018MNRAS.481.1133P, 2020MNRAS.491...51S, 2021JCAP...12..028K, 2021MNRAS.501.1013Z, 2023JCAP...11..043A, 2023PhRvD.107b3530C, 2024JCAP...06..012P, 2024JCAP...03..021K} have been shown to map effectively the history of cosmic structure growth. Combining different tracers of LSS helps overcome the limitations of individual probes, including observational systematics, limited ability to trace the redshift evolution of cosmological parameters, redshifts uncertainties, and complex relationship between galaxy and matter overdensities.

In the past decade, radio continuum surveys have emerged as attractive probes of the LSS. Their ability to survey large areas of the sky and their insensitivity to dust extinction make them valuable for recovering the clustering of galaxies on gigaparsec scales. The radio continuum samples predominantly consists of active galactic nuclei (AGNs) which exhibit a range of physical origins and classifications \citep{1974MNRAS.167P..31F, 1993ARA&A..31..473A, 1995PASP..107..803U, 2012MNRAS.421.1569B, 2014ARA&A..52..589H}, as well as star-forming galaxies \citep[SFGs; e.g.][]{1992ARA&A..30..575C, 2003ApJ...586..794B, 2017MNRAS.466.2312D}. Studying AGNs and SFGs provides insight into the key physical processes that govern galaxy formation and evolution. Their populations have been characterized at very faint flux densities in recent works such as \cite{2017A&A...602A...2S, 2020ApJ...903..139A, 2022MNRAS.516..245W, 2023MNRAS.523.1729B, 2024MNRAS.527.3231W}. Owing to the wide sky coverage achievable with radio surveys and the diversity of radio galaxy populations, these datasets can be used to study relativistic lightcone effects and probe primordial non-Gaussianity \citep{2014MNRAS.442.2511F, 2015ApJ...814..145A, 2015ApJ...812L..22F, 2019JCAP...09..025B, 2020MNRAS.492.1513G}.

A number of past radio continuum surveys such as the Faint Images of the Radio Sky at Twenty centimetres \citep[FIRST;][]{1995ApJ...450..559B}, the NRAO VLA Sky Survey \citep[NVSS;][]{1998AJ....115.1693C}, the Sydney University Molonglo Sky Survey \citep[SUMSS;][]{2003MNRAS.342.1117M}, the TIFR GMRT Sky Survey \citep[TGSS;][]{2017A&A...598A..78I}, and the Cosmic Evolution Survey at 3GHz \citep[COSMOS $3$GHz;][]{2017A&A...602A...1S} have been used to constrain the clustering properties of radio source populations \citep{2004MNRAS.347..787B, 2014MNRAS.440.1527L, 2015ApJ...812...85N, 2018MNRAS.474.4133H, 2019MNRAS.485.5891R}. The current generation of radio surveys conducted with telescopes such as the Australian Square Kilometre Array Pathfinder \citep[ASKAP;][]{2007PASA...24..174J, 2021PASA...38....9H}, the Meer Karoo Array Telescope \citep[MeerKAT;][]{Jonas2009, 2016mks..confE...1J}, and the LOw Frequency ARray (LOFAR, \citealt{2013A&A...556A...2V}) are advancing the field through deeper, high resolution observations over large sky areas of the sky with improved sensitivity \citep{2025PASA...42...88Y, 2025A&A...698A..58Z, 2025A&A...698A.148P, 2025A&A...700A.250L, 2025MNRAS.544.1323H, 2025MNRAS.544.3856S, 2026Ap&SS.371...16C, 2026PASA...43...65R, 2026MNRAS.548ag531P}. In this context, several forecasts have also been developed for the Square Kilometre Array (SKA) Observatory, highlighting the cosmological potential of future radio observations \citep{2012MNRAS.424..801R, 2015aska.confE..18J, 2015aska.confE..25C, 2020PASA...37....7S}.

Any study involving tracers of LSS requires precise knowledge of the redshift distribution of sources. The dominant emission mechanism for radio continuum is synchrotron radiation, which arises from the acceleration of relativistic electrons in magnetic fields. This process produces a nearly featureless power-law spectrum \citep{1992ARA&A..30..575C}, making redshift estimation for radio continuum sources extremely challenging. This difficulty presents two major obstacles for cosmological analyses using radio continuum surveys: (1) accurately characterizing the redshift distribution of the sample, and (2) modelling the evolution of key properties such as galaxy bias and the fractional contributions of different radio source populations. In this paper, we address the first challenge and present a detailed modelling of the redshift distribution for sources detected in the Evolutionary Map of the Universe Pilot Survey 1 (EMU-PS1; \citealt{2021PASA...38...46N}).

In previous radio clustering analyses, redshift distribution information for radio sources has been incorporated in various ways. Studies using the NVSS catalogue \citep{2013arXiv1312.0530M, 2015ApJ...812...85N, 2016A&A...591A.135C} relied on redshift distributions derived from the Combined EIS-NVSS Survey Of Radio Sources (CENSORS) survey \citep{2003MNRAS.346..627B, 2006MNRAS.366.1265B, 2008MNRAS.385.1297B} and Hercules survey \citep{2001MNRAS.328..882W}. However, these surveys include only a small number of sources with known redshifts (approximately 150 for CENSORS and 47 for HERCULES) covering limited sky areas and featuring high flux thresholds ($7.8$ mJy for CENSORS and $1$ mJy for HERCULES at $1.4$ GHz). Cross-matching radio continuum catalogues with optical and near-infrared surveys has become a widely used method to identify redshifts of radio sources \citep{2002IAUS..199...11S, 2019A&A...622A...2W, 2019A&A...622A...3D, 2023ApJ...943..116T}. For example, \cite{2024PASA...41...27G} cross-matched EMU-PS1 catalogue with Dark Energy Survey \citep[DES;][]{2021ApJS..255...20A}, Legacy Imaging Survey \citep{2019ApJS..242....8Z} and WISE $\times$ SuperCosmos \citep{2016ApJS..225....5B} photometric catalogues, obtaining photometric redshifts for $\sim 36\%$ of sources. Recent analyses with the LOFAR Two-meter Sky Survey (LoTSS) Data Release 2 \citep{2024MNRAS.527.6540H, 2024A&A...681A.105N} have utilized deep field observations \citep{2021A&A...648A...1T, 2021A&A...648A...2S, 2021A&A...648A...3K} to calibrate redshift distributions and \cite{2024MNRAS.530.2994T} measured the unresolved radio source background. In a recent study, \cite{2025PASA...42...77P} adopted the Galaxy and Mass Assembly (GAMA; \citealt{2011MNRAS.413..971D,2020MNRAS.496.3235B}) and COSMOS redshift distributions to infer the luminosity functions of {EMU early science radio sources \citep{2022MNRAS.512.6104G}.

In addition to the aforementioned strategies, there are available two extragalactic radio continuum simulations: the Tiered Radio Extragalactic Continuum Simulation (hereafter \texttt{TRECS}; \citealt{2019MNRAS.482....2B}) and the European SKA Design Study (hereafter \texttt{SKADS}) Simulated Skies \citep{2008MNRAS.388.1335W}. Many studies have utilised the redshift distributions from these simulations to derive cosmological inferences from the clustering of radio sources \citep{2019A&A...623A.148D, 2019MNRAS.488.5420H, 2022MNRAS.517.3785B, 2023A&A...671A..42P, 2024PASA...41...70V}. Recently, \citealt{2025PASA...42...62T} (hereafter, T25) used these simulated redshift distributions to constrain $\sigma_{8}-$parameter via the cross-correlation between EMU-PS1 radio catalogue and \textit{Planck} CMB lensing convergence map.

As we reach the time of precision cosmology with radio continuum data, an accurate understanding of the radio source population is necessary; otherwise a systematic problem with the $n(z)$ may lead to an offset in the probability distribution in the cosmological parameters. One option is direct optical measurement of redshifts of a positional cross-matched sample. For example, in small fields such as VLA-COSMOS and MIGHTEE data the optical identification fraction is $>$90\% \citep{2020ApJ...903..139A,2024MNRAS.527.3231W}.  The LOFAR Two-metre Sky Survey (LoTSS)  deep fields are similarly complete \citep{2021A&A...648A...4D, 2024A&A...692A...2B}, although their photo-$z$s may not be as accurate as can be achieved in COSMOS. Here we offer an alternative approach to direct matching  for learning about $n(z)$, which should provide complimentary and consistent results, which uses data over the entire survey area, and while it does not provide individual source redshifts, should be perfectly representative.

As in the case of optical surveys, redshift tomography can also be employed to infer the clustering properties of the radio sample and characterise its redshift distribution \citep{2013arXiv1303.4722M}. Our focus in this work is to constrain the EMU-PS1 redshift distribution through tomographic cross-correlation measurements with the Dark Energy Survey (DES) magnitude-limited (MagLim) galaxy sample \citep{2021PhRvD.103d3503P, 2022PhRvD.106j3530P}. We adopt a parametric model for the EMU-PS1 redshift distribution previously used by \cite{2024A&A...681A.105N} in their LoTSS DR2 analysis. Our analysis derives redshift constraints exclusively from angular cross-power spectra between EMU-PS1 and DES galaxies, divided into six redshift bins. We do not incorporate any secondary redshifts information from cross-matching or deep field observations. This paper serves to demonstrate the feasibility of estimating redshift distributions directly from large-area radio continuum surveys such like EMU main survey \citep{2025PASA...42...71H} and forthcoming SKA.\\

The paper is organised as follows: Section \ref{sec:theory} introduces the theoretical background for clustering statistics and the angular power spectrum model. Section \ref{sec:data} describes the EMU-PS1 and DES datasets, while Section \ref{sec:methodology} outlines the assumptions and methodology used to analyse them. The measurements of angular power spectra and constraints on EMU-PS1 redshift distribution are presented in Section \ref{sec:results}. We discuss and summarise our findings in Section \ref{sec:conclusion}.

\section{Theory} \label{sec:theory}
The main probes studied in this paper are the projected overdensities of galaxies ($\delta_{g}$) in the EMU and DES samples. The projected overdensities quantifies fluctuations around the average galaxy number density $\overline{N}_{g}$ as
\begin{equation}
    \delta_{g}(\hat{\bm{n}}) = \frac{N_{g}(\hat{\bm{n}})-\overline{N}_{g}}{\overline{N}_{g}},
    \label{eq:gal_overdensity_theory}
\end{equation}
where $N_{g}(\hat{\bm{n}})$ is the number of galaxies at position $\hat{\bm{n}}$. The projected overdensity is related to the three-dimensional galaxy overdensity $\delta_{g}(\chi\hat{\bm{n}},z)$ with the relation
\begin{equation}
    \delta_{g}(\hat{\bm{n}}) = \int \de z\,n(z)\varDelta_{g}(\chi\hat{\bm{n}},z)
    \label{eq:galaxy_overdensity_2d_3d_relation}
\end{equation}
where $n(z)\equiv\de N/\de z$ is the normalised redshift distribution of galaxies, and $\chi(z)$ is the comoving distance to redshift $z$, such that $\de\chi/\de z=c/H(z)$, with $H$ the Hubble rate.

The projected galaxy overdensity field $\delta_{g}$ can be decomposed in terms of spherical harmonic coefficients $g_{\ell m}$. The angular power spectrum between the EMU and DES galaxy samples can then defined as (assuming statistical isotropy)
\begin{equation}
    \langle g_{\ell m}^{A}\,g_{\ell' m'}^{B} \rangle = \delta^{K}_{\ell\ell'}\,\delta^{K}_{mm'}\,C_{\ell}^{AB}\;,
    \label{eq:angular_power_spectrum}
\end{equation}
where $A,B\in \{\text{EMU},\text{DES}\}$ and $\delta^{K}$ is the Kronecker delta function.

The angular power spectrum between two projected galaxy fields $A$ and $B$ can be modelled with the knowledge of two key components. First is the matter auto-power spectrum, $P_{mm}(k,z)$, which describes the clustering of the underlying matter density field. Second a prescription of galaxy bias which relates the luminous tracers of the LSS to the underlying total matter distribution. For the range of scales $k<0.2\,h/\text{Mpc}$ probed in this study, we adopt a linear, scale independent galaxy bias for both EMU and DES fields. Under these assumptions, the matter power spectrum can be factorised into redshift and scale-dependent terms as
\begin{equation}
    P_{gg}^{AB}(k,z) = b^{A}(z)\,b^{B}(z)\,P_{mm}(k,z)\;,
    \label{eq:galaxy_to_matter_power_spectra}
\end{equation}
where $b(z)$ is the galaxy bias, and $P_{mm}(k,z)$ is the matter power spectrum at redshift $z$ computed using \texttt{CAMB} \citep{2000ApJ...538..473L} including the \texttt{HALOFIT} prescription \citep{2003MNRAS.341.1311S} for non-linear evolution. The model angular power spectrum can, then, be written as
\begin{equation}
    C_{\ell}^{AB} = \frac{2}{\pi}\int \de k\, k^{2}\,W_{\ell}^{A}(k)\,W_{\ell}^{B}(k)\,P_{mm}(k)\;.
    \label{eq:theory_cls}
\end{equation}
$W_{\ell}(k)$ is the window function given by
\begin{equation}
    W_{\ell}(k) = W_{\ell}^{\delta}(k) + W_{\ell}^{\mu}(k)
    \label{eq:galaxy_window_function}
\end{equation}
where $W_{\ell}^{\delta}(k)$ is contribution from galaxy density
\begin{equation}
    W_{\ell}^{\delta}(k) = \int\de z\,b(z)\,n(z)\,j_{\ell}[k\,\chi(z)]\;,
    \label{eq:galaxy_density_window_function}
\end{equation}
and $W_{\ell}^{\mu}(k)$ is the contribution from magnification of galaxies
\begin{equation}
\begin{split}
    W_{\ell}^{\mu}(k) = \frac{3\,\Omega_{m,0}\,H_{0}}{2\,c}\,C^{\mu}& \int\de z\,(1+z)\,\frac{\chi(z)}{H(z)}\,j_{\ell}[k\chi(z)]\\
    \times & \int\limits_{z}^{z_{H}}\de z'\,n(z')\,\bigg[\frac{\chi'(z)-\chi(z)}{\chi'(z)}\bigg]\;,
\end{split}
    \label{eq:galaxy_mag_window_function}
\end{equation}
with $j_{\ell}(k\,\chi)$ being spherical Bessel functions, $c$ is the speed of light, $z_{H}$ is redshift to the horizon and $H(z)$ is the Hubble parameter. $\Omega_{m,0}$ is the present day value of matter density parameter and $H_{0}$ is the Hubble constant. $C^{\mu}$ is the constant magnification bias amplitude \citep{2022PhRvD.106j3530P, 2023MNRAS.523.3649E}. The magnification component for EMU-PS1 cannot be estimated directly due to the lack of redshifts for radio sources. Recently, \cite{2024PASA...41...27G} cross-matched EMU-PS1 with optical and infrared galaxy catalogues, finding photometric redshifts for $36\%$ of the radio sources. We used this value-added catalogue to estimate EMU-PS1 magnification bias amplitude, finding values of $C^{\mu}=0.28$, $0.13$, $0.03$, $-0.01$, $-0.07$ and $-0.02$ across the six redshift bins, respectively. The small values of $C^{\mu}$ modify the model angular power spectra by less than $4\%$. Hence, we neglect the EMU-PS1 magnification component in this study. With the ongoing EMU main survey (see \citealt{2025PASA...42...71H} for details), improved constraints on magnification bias are expected through cross-matching with galaxy surveys such as DES, \textit{Euclid} \citep{2025A&A...697A...1E} and WISE \citep{2010AJ....140.1868W}. The values of magnification bias amplitude for DES tomographic bins are provided in Table \ref{tab:phy_prop_emu_des}. The model angular power spectra (Eq.\ \ref{eq:theory_cls}) are computed following the method described in \cite{2020JCAP...05..010F}.

\begin{table}[hbt!]
\begin{threeparttable}
\caption{Physical properties of EMU-PS1 and DES fields. $N_{g}$ is the total number of sources, $\overline{n}_{g}$ is the density of sources and $C^{\mu}$ is the amplitude of magnification bias.}
\label{tab:phy_prop_emu_des}
\begin{tabular}{cccccc}
\toprule
\headrow & Area\tnote{a} & $z-$range & $N_{g}$ & $\overline{n}_{g}$\tnote{b} & $C^{\mu}$\\
\headrow & & $[z_{\text{min}},z_{\text{max}})$ & & &\\
\midrule
EMU-PS1 & $270$ & - & $184,203$ & $0.189$ & -\\
\midrule
        &  & [$0.20,\,0.40$) & $2,236,473$ & $0.150$ & $0.43$\\
        &  & [$0.40,\,0.55$) & $1,599,500$ & $0.107$ & $0.30$\\
DES     & $5000$ & [$0.55,\,0.70$) & $1,627,413$ & $0.109$ & $1.75$\\
        &  & [$0.70,\,0.85$) & $2,175,184$ & $0.146$ & $1.94$\\
        &  & [$0.85,\,0.95$) & $1,583,686$ & $0.106$ & $1.56$\\
        &  & [$0.95,\,1.05$) & $1,494,250$ & $0.100$ & $2.96$\\
\bottomrule
\end{tabular}
\begin{tablenotes}[hang]
\item[a] in deg$^{2}$
\item[b] gal arcmin$^{-2}$
\end{tablenotes}
\end{threeparttable}
\end{table}

In this paper, we adopt the flat $\Lambda$CDM cosmology with best fitting \textit{Planck} + WP + lensing cosmological parameters, as described in \cite{2020A&A...641A...6P}. Here, WP refers to WMAP polarization data at low multipoles \citep{2013ApJS..208...20B}, and lensing refers to the inclusion of \textit{Planck} CMB lensing data in the parameter likelihood \citep{2020A&A...641A...8P}. The list of cosmological parameters quoted in Table \ref{tab:fixed_cosmo_params} are fixed in our analysis.
\begin{table}[hbt!]
\begin{threeparttable}
\caption{Cosmological parameters assumed fixed in this work, taken from \cite{2020A&A...641A...6P}}
\label{tab:fixed_cosmo_params}
\begin{tabular}{wc{1.3cm}wc{1.3cm}wc{1.3cm}wc{1.3cm}wc{1.3cm}}
\toprule
\headrow $\Omega_{c,0}$ & $\Omega_{b,0}$  & $H_{0}$\tnote{a} & $\sigma_{8}$ & $n_{s}$\\
\midrule
$0.265$ & $0.049$ & $67.32$ & $0.811$ & $0.9645$\\ 
\bottomrule
\end{tabular}
\begin{tablenotes}[hang]
\item[a] in km s$^{-1}$ Mpc$^{-1}$
\end{tablenotes}
\end{threeparttable}
\end{table}

\section{Data}\label{sec:data}

\subsection{EMU Pilot Survey 1}

The Evolutionary Map of the Universe \citep[EMU;][]{2011PASA...28..215N,2025PASA...42...71H}\footnote{ EMU project page: \url{https://emu-survey.org/}} is an ongoing radio continuum survey conducted with the ASKAP radio telescope \citep{2021PASA...38....9H}. EMU aims to survey the entire southern hemisphere within five years (expected completion in 2028), cataloguing approximately 20 million extragalactic sources. In this study, we use data from the EMU Pilot Survey 1 (EMU-PS1; \citealt{2021PASA...38...46N}) which covers $\sim 270$ deg$^{2}$ observed at $943$ MHz, with an angular resolution of $11-18$ arcsec and a rms depth of $25-30\,\mu$Jy/beam.

A catalogue of radio sources for the EMU-PS1 was generated using two different source finding algorithms, \texttt{Selavy} \citep{Whiting_Humphreys_2012, 2017ASPC..512..431W} and \texttt{PyBDSF} \citep{2015ascl.soft02007M}. The official \texttt{Selavy} source and components catalogues were released as part of the pilot survey. Additionally, we applied the \texttt{PyBDSF} source finder to the EMU-PS1 image for our companion study in T25. The total number of sources in the catalogue is sensitive to the choice of source-finding algorithm, and we refer the readers to T25 for a detailed comparison of \texttt{Selavy} and \texttt{PyBDSF}, including their impact on source counts. For the angular scales used in this analysis ($\ell < 250$), we find that the EMU-PS1 galaxy auto-power spectra derived from the \texttt{Selavy} and \texttt{PyBDSF} catalogues are in excellent agreement (see Figure 3 of T25). Accordingly, for the present work, we use the \texttt{PyBDSF}-generated catalogue, which contains a total of 224,100 sources. To ensure a conservative sample, we apply a flux density cuts at $180\,\mu$Jy and $400\,\mu$Jy (approximately a $6\sigma$ and $\sim 10\sigma$ detection threshold), resulting in a final catalogue of $184,203$ and $85,895$ radio sources, respectively.
\subsection{Dark Energy Survey}

\begin{figure}[hbt!]
\centering
\includegraphics[width=\linewidth]{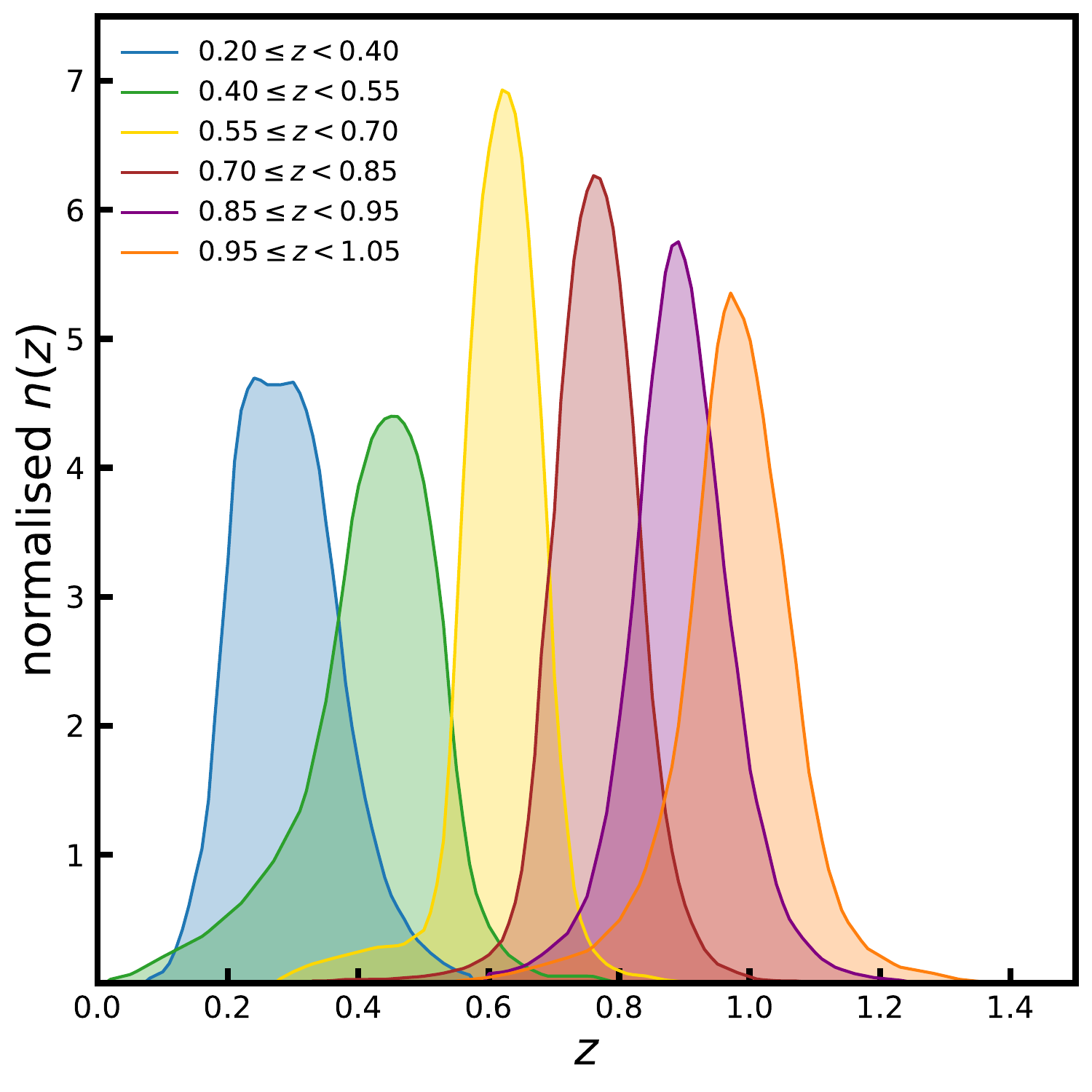}
\caption{Redshift distribution of six DES tomographic bins computed using DNF algorithm and calibrated using clustering redshifts}
\label{fig:gal_dist_des_maglim}
\end{figure}

The Dark Energy Survey (DES)\footnote{DES project page: \url{https://www.darkenergysurvey.org/}} was a six-year observing programme conducted with the Dark Energy Camera \citep{2015AJ....150..150F} mounted on the Blanco $4$-meter telescope at the Cerro Tololo Inter-American Observatory (CTIO) in Chile. The survey covered $5000$ deg$^{2}$ of the southern sky in $grizY$ filters, including the EMU-PS1 footprint. In this study, we utilise the magnitude-limited (MagLim) galaxy sample \citep{2021PhRvD.103d3503P, 2022PhRvD.106j3530P} constructed from the first three years of DES observations \citep{2018ApJS..239...18A}. The MagLim sample is defined with a cut on the $i-$band magnitude that depends linearly on photometric redshift, $i<4z_{\text{ph}}+18$ \citep{2022PhRvD.105b3520A}.

The photometric redshifts for MagLim galaxies were estimated using the Directional Neighbourhood Fitting algorithm \citep[DNF;][]{2016MNRAS.459.3078D, 2021ApJS..254...24S}. The DNF algorithm computes a point estimate $z_{\text{DNF}}$ of each galaxy's redshift by fitting through the $80$ nearest neighbours in colour-magnitude space constructed from a spectroscopic reference set. This reference set includes objects from $24$ different spectroscopic catalogues, such as the Sloan Digital Sky Survey (SDSS) DR14 \citep{2018ApJS..235...42A}, the Optical redshifts for the Dark Energy Survey program \citep[OzDES;][]{2020MNRAS.496...19L}, the VIMOS Public Extragalactic Redshift Survey \citep[VIPERS;][]{2014A&A...562A..23G}, and the Physics of the Accelerating Universe photometric catalogue \citep{2019MNRAS.484.4200E}. In addition to point estimates, DNF provides individual redshift posterior derived from the fitting residuals \citep{2022PhRvD.106j3530P}.

The MagLim sample is divided into six tomographic bins with edges at $z =[0.20, 0.40, 0.55, 0.70, 0.85, 0.95, 1.05]$. The number density of galaxies in each of these six bins are provided in Table \ref{tab:phy_prop_emu_des}. The corresponding redshift distributions, validated using galaxy clustering cross-correlations \citep{2022MNRAS.513.5517C}, are shown in Figure \ref{fig:gal_dist_des_maglim}. The DES data used in this analysis are publicly available at \url{https://des.ncsa.illinois.edu/}.

\subsection{\textit{Planck} PR3}

T25 measured the cross-correlation between EMU-PS1 and \textit{Planck} CMB lensing convergence to estimate the EMU-PS1 galaxy bias and $\sigma_{8}$-parameter. In this work we measure the cross-power spectrum between EMU-PS1 radio sources and \textit{Planck} lensing convergence map ($\kappa$) to validate the EMU-PS1 redshift distribution and galaxy bias, obtained from cross-correlating between EMU-PS1 with DES tomographic bins. The $\kappa-$map we use comes from the 2018 \textit{Planck} data release\footnote{\url{https://pla.esac.esa.int/\#cosmology}} described in \cite{Planck2020VIII}. It uses the SMICA DX12 CMB maps to reconstruct the lensing potential, covering $\sim 67$\% of the sky. We use the $\kappa-$map derived from a minimum-variance estimate of temperature and polarization data. The spherical harmonic coefficients for lensing convergence are provided by \textit{Planck} data package in HEALPix\footnote{\url{https://healpix.jpl.nasa.gov/}} \citep{2005ApJ...622..759G} format with $N_{\text{side}}=4096$. We downgrade these coefficients to $N_{\text{side}}=512$ for cross-correlation with EMU-PS1 data.

\section{Methodology}\label{sec:methodology}

\begin{figure*}[hbt!]
    \centering
    \includegraphics[width=\linewidth]{Figures/maps_compressed.pdf}
    \caption{Galaxy overdensity maps used in the study. (a) \textit{Planck} CMB lensing convergence, (b) EMU-PS1 weighted overdensity map following Eq. \ref{eq:gal_overdensity_data}, (c)-(h) DES galaxy overdensity maps in six tomographic bins. The blue contour in DES maps shows the EMU-PS1 footprint. The colorbars indicate the strength of overdensities (in pixels), with masked region shown in grey colour. The grid spacing is $5^{\circ}$ (in both ra and dec) for EMU-PS1, and $30^{\circ}$ for DES.}
    \label{fig:maps_emu_des}
\end{figure*}

\subsection{Power spectra}

The computation of power spectra requires accurately specifying the survey footprint and characterising the fluctuations in observed number of sources due to image noise. The observed galaxy overdensity maps in Eq. \ref{eq:gal_overdensity_theory} is modified to
\begin{equation}
    \hat{\delta}_{g}(\hat{\bm{n}}) = \frac{\hat{N}_{g}(\hat{\bm{n}})}{\overline{N}_{g}\,w_{g}(\hat{\bm{n}})} - 1,
    \label{eq:gal_overdensity_data}
\end{equation}
where $w_{g}(\hat{\bm{n}})$ is the weight map which defines the variations in observed number of sources over the survey footprint due to observational noise. $\overline{N}_{g}$ now becomes the weighted mean source density per pixel
\begin{equation}
    \overline{N}_{g} = \frac{\langle \hat{N}_{g}(\hat{\bm{n}})\rangle}{\langle w_{g}(\hat{\bm{n}})\rangle},
    \label{eq:weighted_number_density}
\end{equation}
where $\langle\cdot\rangle$ denotes average over all observed pixels.

The construction of weight maps for EMU-PS1 is described in section 4.1 of T25, while the weights for DES galaxies are available from the public dataset. We build the weighted EMU-PS1 and DES galaxy overdensity maps at \texttt{HEALPix} \citep{2005ApJ...622..759G} resolution $N_{\text{side}} = 512$ using Eq. \ref{eq:gal_overdensity_data}. To avoid heavily masked pixels, we set both $w_{g}(\hat{\bm{n}})$ and $\delta_{g}(\hat{\bm{n}})$ to zero where $w_{g}(\hat{\bm{n}})<0.5$. The \textit{Planck} CMB lensing convergence map and the weighted galaxy overdensity map for EMU-PS1 are shown in Figure \ref{fig:maps_emu_des}a and \ref{fig:maps_emu_des}b, respectively, while Figures \ref{fig:maps_emu_des}c–\ref{fig:maps_emu_des}h display the DES galaxy overdensity maps for six tomographic bins.

The computation of angular power spectra via Eq.~\ref{eq:angular_power_spectrum} assumes full sky coverage. However, in practice we only observe a fraction of sky in galaxy surveys, and can only compute the pseudo power spectra. The incomplete sky coverage induces coupling between different harmonic modes. We used a pseudo-$C_{\ell}$ estimator \citep{1973ApJ...185..413P, 2002ApJ...567....2H} implemented in \texttt{NaMaster} python code \citep{2019MNRAS.484.4127A} to compute the full sky power spectrum from the pseudo power spectrum. The pseudo-$C_{\ell}$ estimator accounts for the harmonic mode coupling by computing the coupling matrix by factoring in the survey area \citep{2002ApJ...567....2H, 2022MNRAS.515.1993S}. Using this framework, we compute the EMU-PS1 galaxy auto-power spectrum ($C_{\ell}^{\text{EMU}}$), DES galaxy auto-power spectrum in six tomographic bins ($C_{\ell}^{\text{DES}}$), the cross-power spectra between EMU-PS1 and DES tomographic bins ($C_{\ell}^{\text{EMU}\times\text{DES}}$), and the cross-power spectrum between EMU-PS1 and \textit{Planck} CMB lensing convergence ($C_{\ell}^{\text{EMU}\times\kappa}$).

\subsection{EMU source redshift distribution}
The theoretical angular power spectra in Eq. \ref{eq:theory_cls} requires knowledge of the redshift distribution of EMU-PS1 sources. As described in Section \ref{sec:introduction}, in absence of any direct measurements of redshifts for radio continuum sources, many studies with clustering of radio sources relies heavily on \texttt{TRECS} and \texttt{SKADS} simulations. These simulations include contributions from AGNs and SFGs as the main tracers of radio galaxy populations. Recently, T25 used these simulated redshift distributions to model EMU-PS1 auto-power spectrum and its cross-correlation with CMB lensing convergence.

In this work, we attempt to constrain the redshift distribution of EMU-PS1 sources based on the measured cross-correlation with DES tomographic bins. We examine the efficacy of \texttt{TRECS} to model the cross-power spectra. We also consider the parametric model from \cite{2024A&A...681A.105N} for the EMU-PS1 $n(z)$, to test departures from \texttt{TRECS}. 
The model redshift distribution has three free parameters and is given by
\begin{equation}
    n(z) \propto \frac{z^2}{1 + z}\,\left[\exp\left(-\frac{z}{z_0}\right) + \frac{r^2}{\left(1 + z\right)^\gamma}\right]\;
    \label{eq:model_nz_lotssdr2}
\end{equation}
The redshift distribution model is physically motivated to include contributions from both AGN and SFG populations in the EMU-PS1 redshift distribution. The parameter $z_{0}$ primarily controls the peak of the redshift distribution, $r$ parametrises the relative fraction of AGN and SFG contributions, and $\gamma$ affects the decay rate of the high redshift tail.

\subsection{Galaxy bias model}\label{sec:galaxy_bias_model}
The final ingredient required for our analysis is a model of the linear galaxy bias from EMU-PS1 and DES tomographic bins. Following \cite{2022PhRvD.105b3520A} and \cite{2022PhRvD.106j3530P}, we adopt a linear constant galaxy bias model for DES tomographic bins, $b^{i}(z) = b_{g}^{i}$. For EMU-PS1, different galaxy bias models including constant galaxy bias, constant amplitude galaxy bias, and quadratic galaxy bias were studied in our companion T25 paper. When testing the efficacy of \texttt{TRECS} redshift distribution in this work, we use a mixture model for the evolution of EMU PS1 galaxy bias given by \citep{2014MNRAS.442.2511F}
\begin{equation}
b_{\text{eff}}(z) = \frac{b_{\text{AGN}}(z)\, n_{\text{AGN}}(z)+b_{\text{SFG}}(z)\, n_{\text{SFG}}(z)}{n_{\text{AGN}}(z)+n_{\text{SFG}}}
\label{eq:effective_bias}
\end{equation}
where $n_{\text{AGN}}(z)$ and $n_{\text{SFG}}(z)$ are AGN and SFG fraction, respectively, derived from \texttt{TRECS} simulations. We adopt a constant amplitude galaxy bias model for SFG, $b_{\text{SFG}}(z) = b_{\text{SFG},0}/D(z)$, and polynomials up-to degree 3 to model AGN galaxy bias.

However, for analyses with the parametric model (Eq.\ \ref{eq:model_nz_lotssdr2}) we adopt a constant amplitude galaxy bias and polynomials up-to degree two, to model the evolution of EMU-PS1 effective galaxy bias. Furthermore, the free parameter $r$ in Eq\,\ref{eq:model_nz_lotssdr2} takes values between $r\in[0,\infty)$, which can lead to runaway chains in MCMC sampling of the parameter space. To avoid such conditions, we reparametrise $r^{2} \to (1-y)/y$ and sample $y$ between $[0,1]$ for the mathematical convenience of exploring the parameter space. We note that our new parameter $y$ does not denote the ratio of AGNs to SFGs, instead it is a mere mathematical transformation.

\subsection{Covariance matrix}
A robust estimate of covariance matrix for the power spectra is crucial to estimate parameters. We used the pseudo-power spectra measured from data after rescaling with the survey sky fraction ($f_{\mathrm{sky}}^{\mathrm{EMU-PS1}}=0.0065$ and $f_{\mathrm{sky}}^{\mathrm{DES}}=0.1187$) to generate mock EMU-PS1 and DES tomographic density fields using \texttt{GLASS} \citep{2023OJAp....6E..11T}. We applied the appropriate angular survey selection function to mock data. To add noise to these simulated density fields, we generated number count maps where the value in each pixel is drawn from a Poisson distribution with mean $\lambda$ given by
\begin{equation}
	\lambda(\hat{\bm{n}}) = \overline{n}(1+g(\hat{\bm{n}}))
	\label{eq:Poisson_noise}
\end{equation}
where $\overline{n}$ is the mean number of sources per pixel (measured from observed EMU-PS1 and DES source counts) and $g(\hat{\textbf{n}})$ is the corresponding simulated galaxy over-density map. Finally, we compute the simulated power spectra and bin them into band-powers similar to observed data. We create $1000$ mock realisations and compute the sample covariance matrix using the relation
\begin{equation}
    {\sf K}_{\ell\ell'}^{AB,CD} = \frac{1}{{N}_{{s}}-1}\sum_{i=1}^{{N}_{{s}}}\left(\tilde{C}^{AB,i}_\ell-\left\langle \tilde{C}_{\ell}^{AB} \right \rangle\right)\left(\tilde{C}^{CD,i}_\ell-\left\langle  \tilde{C}_{\ell}^{CD}\right\rangle\right),
    \label{eq:sample_covariance}
\end{equation}
where $A,B,C,D \in \{g^{\text{DES}},g^{\text{EMU}}\}$, ${N}_{{s}}$ is the total number of simulations, $\tilde{C}^{gg,i}_\ell$ is the power spectrum estimated from the $i\text{th}$ simulation and
\begin{equation}\label{eq:mock_avg}
    \left\langle  \tilde{C}_{\ell}^{gg}\right\rangle = \frac{1}{{N}_{{s}}}\sum_{i=1}^{{N}_{{s}}}\tilde{C}^{gg,i}_\ell
\end{equation}
The method of computing covariance matrix allows us to include properties of EMU-PS1 field without any assumptions about its source redshift distribution or galaxy bias. We account for the Anderson-Hartlap correction factor when computing the inverse of the sample covariance matrix \citep{Anderson2004,2007A&A...464..399H}
\begin{equation}
    {\sf K}_{\ell\ell'}^{-1} \to \frac{{N}_{{s}}-{N}_{{d}}-2}{{N}_{{s}}-2}\,{\sf K}_{\ell\ell'}^{-1}
\end{equation}
where ${N}_{{d}}$ is the size of the data vector.

\subsection{Parameter estimation}\label{sec:parameter_estimation}
As first step, we used maximum likelihood estimation with $C_{\ell}^{\text{DES}}$ to estimate the galaxy bias in independent DES tomographic bins. This enabled us to validate our analysis pipeline by comparing our estimates with previous DES galaxy bias measurements.  We, then, used the joint data vector of six cross-power spectra $\bm{d}_{\ell} = \{C_{\ell}^{\text{EMU}\times\text{DES}}\}$ to constrain EMU-PS1 $n(z)$ and galaxy bias evolution. The log-likelihood function is given 
\begin{equation}
    \log \mathcal{L} = -\frac{1}{2}[\bm{d}_{\ell}-\bm{t}_{\ell}(\theta)]^\top\,({\sf K}^{-1})_{\ell\ell'}\,[\bm{d}_{\ell}-\bm{t}_{\ell}(\theta)],
\label{eq:joint_likeli}
\end{equation}
where $\bm{t}_{\ell}(\theta)$ is the theory vector. $\theta$ represents the set of free parameters; $\theta=\{b_{\text{SFG},0}, b_{\text{AGN},0}, b_{\text{AGN},1}, b_{\text{AGN},2}, b_{\text{AGN},3}\}$ for analysis using \texttt{TRECS} $n(z)$, and $\theta=\{b_{0}, b_{1}, b_{2}, z_{0}, y, \gamma\}$ for model $n(z)$ given by Eq. \ref{eq:model_nz_lotssdr2}. ${\sf K}_{\ell\ell'}$ is the joint covariance matrix given by Eq.\ \ref{eq:full_cov_matrix}.

The parameter priors used in our analysis are mentioned in Table \ref{tab:parameter_priors}. We used the \texttt{PyMC} package \citep{pymc2023} to effectively sample the parameter space and generate posterior distributions. We put limits on model $n(z)$ such that its tail drops to zero at $z=6$. Additionally, we check the first and second derivatives of our model $n(z)$ to avoid tophat-like distributions of EMU-PS1 sources. With the best-fit values of parameters obtained using \texttt{PyMC}, we fit the EMU-PS1 auto-power spectrum ($C_{\ell}^{\text{EMU}}$) and its cross-power spectrum with \textit{Planck} CMB lensing convergence ($C_{\ell}^{\text{EMU}\times\kappa}$). These two power-spectra serve as validation for the best-fit EMU-PS1 $n(z)$.

\begin{table}[hbt!]
\begin{threeparttable}
\caption{The parameters and their priors used in the analysis. $b_{i}^{\text{DES}}$ is the galaxy bias for the $i$th DES tomographic bin. $b_{\text{SFG},0}$ and $b_{\text{AGN},i}$ is the galaxy bias for EMU-PS1 radio sources when using \texttt{TRECS} $n(z)$. $b_{i}$ is the EMU-PS1 galaxy bias for analysis with model $n(z)$. The parameters $z_{0}, w,\text{ and }\gamma$ control the model $n(z)$ in Eq.\ \ref{eq:model_nz_lotssdr2}.}
\label{tab:parameter_priors}
\begin{tabular}{wc{2.6cm}wc{3.9cm}}
\toprule
\headrow Parameter & Prior\\
\midrule
$b_{i}^{\text{DES}}$ ($i\in {1,\cdots,6}$) & Flat$(0,5)$\\ 
$b_{\text{SFG},0}$ & LogNormal$(0.5,0.5)$\\ 
$b_{\text{AGN},i}$ ($i\in {0,\cdots,3}$) & LogNormal$(0.5,0.5)$\\
$b_{i}$ ($i\in {0,\cdots,2}$) & LogNormal$(0.5,0.5)$\\
$z_{0}$ & LogNormal$(0.0,1.5)$\\
$y$ & Beta$(1.5,1.5)$\\
$\gamma$ & LogNormal$(1.5,0.5)$\\
\bottomrule
\end{tabular}
\end{threeparttable}
\end{table}

\section{Pipeline validation}\label{sec:simulations}

\begin{figure*}[!hbt]
\begin{subfigure}{0.5\linewidth}
    \centering
    \includegraphics[width=\linewidth]{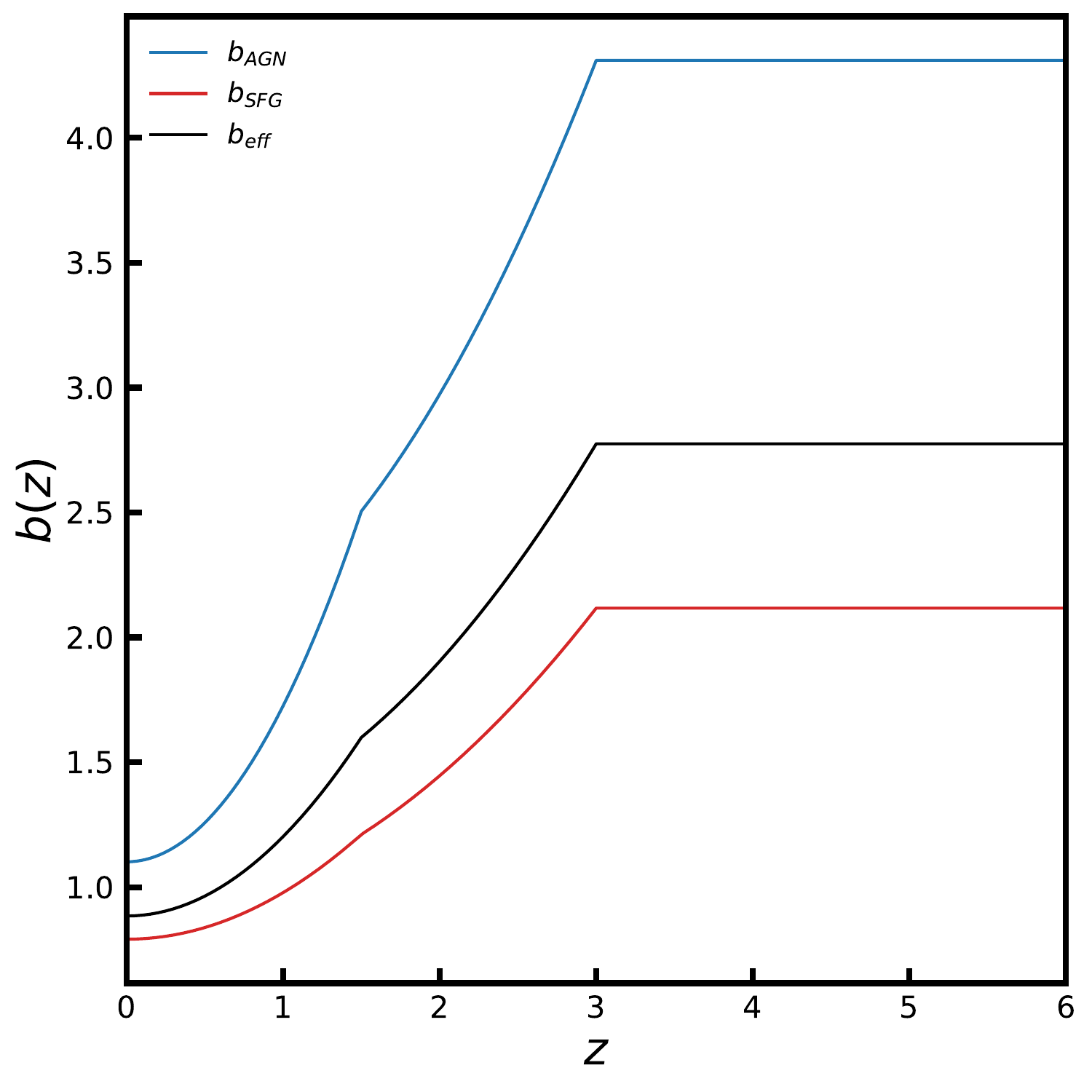}
    \caption{}
    \label{fig:gal_bias_agn_sfg_trecs_normal}
\end{subfigure}%
\begin{subfigure}{0.5\linewidth}
    \centering
    \includegraphics[width=\linewidth]{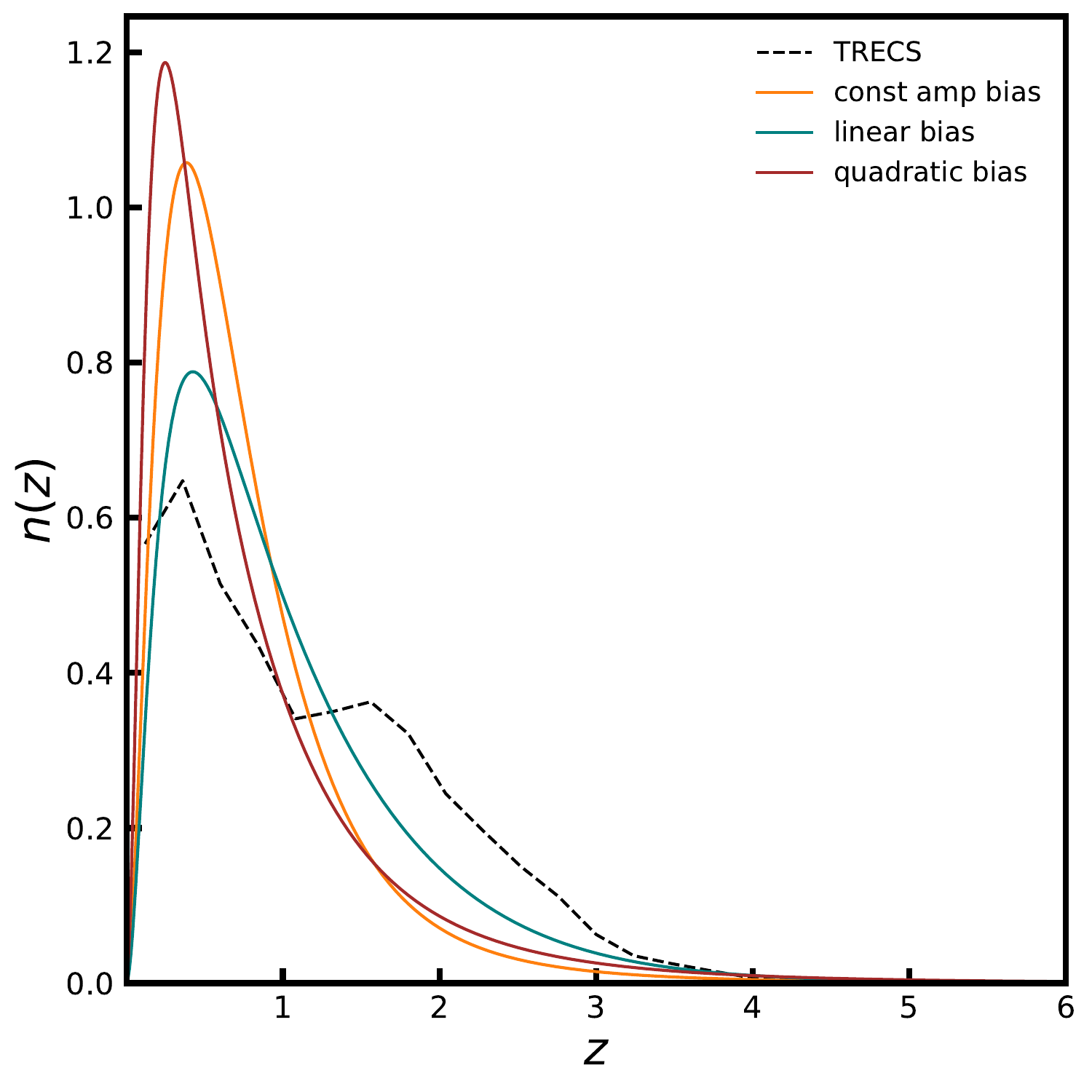}
    \caption{}
    \label{fig:recovered_nz_trecs_normal}
\end{subfigure}
    \caption{Pipeline validation results for cross-correlation approach to estimating radio galaxy bias and redshift distribution. (a) The fiducial galaxy bias model used for creating mock EMU-PS1 catalogues. The galaxy bias for AGNs is shown in blue, for SFGs in red, and the effective galaxy bias for their joint population is shown in black. (b) Best-fit EMU-PS1 redshift distributions recovered from simulations with different assumptions for galaxy bias model used in our baseline analysis, i.e. constant amplitude bias (in orange), linear bias (in green). The black dashed line shows the fiducial redshift distribution of \texttt{TRECS} AGN and SFG joint sample. The redshift distributions are individually normalised to a unit area.}
    \label{fig:pipeline_validation}
\end{figure*}

Before analysing data, it is essential to assess how well our parametric redshift distribution model (Eq. \ref{eq:model_nz_lotssdr2}) and analysis choices (described in Section \ref{sec:methodology}) can recover the underlying true redshift distribution of EMU-PS1 radio sources. In this section, we test the robustness of our analysis pipeline against internal systematics. We used \texttt{GLASS} to generate $100$ mock simulations of correlated EMU-PS1 and the six DES tomographic galaxy fields. The mocks adopt the same angular coverage and number density of sources as described in Table \ref{tab:phy_prop_emu_des}.

To simulate photometric redshift errors for DES, we randomise the simulated redshifts assuming a Gaussian error distribution with width obtained from the DNF estimates of observed DES galaxies. For EMU-PS1, we considered AGNs and SFGs as independent populations and used \texttt{TRECS} simulations to obtain their fractional contributions as well as redshift distributions. The \texttt{TRECS} simulation does not provide fluxes directly at the EMU observation frequency of $944$ MHz. To address this, we interpolate the available fluxes between the provided frequency bands to estimate the flux at 944 MHz and then apply a flux threshold of $180\,\mu$Jy.} However, for the analysis we considered the joint simulated sample of AGNs and SFGs as a single population. 

The prescriptions for AGN and SFG galaxy bias in the simulations is based on different population types (i.e. normal star-forming, starburst, radio-quiet quasar, FRI and FRII radio galaxies) following \cite{2012MNRAS.424..801R}. The galaxy bias beyond redshift $z\sim 3$ assumes a constant value to avoid excessive clustering strength at high redshifts. We emphasise that the galaxy bias models for AGNs and SFGs are intended to only probe potential systematic effects in our methodology for estimating the EMU-PS1 redshift distribution. The galaxy biases for AGNs, SFGs and their joint sample are shown in Figure \ref{fig:gal_bias_agn_sfg_trecs_normal}. The $b_{\text{eff}}(z)$ represents the evolution of galaxy bias for the joint sample computed using Eq.\ \ref{eq:effective_bias}.

To estimate parameters, we apply the same parametric redshift distribution (Eq. \ref{eq:model_nz_lotssdr2}) and galaxy bias models (Section \ref{sec:galaxy_bias_model}) used for the real data. Figure \ref{fig:recovered_nz_trecs_normal} shows the recovered EMU-PS1 redshift distributions with different assumptions of galaxy bias models. The peak location of recovered redshift distributions is consistent with the fiducial \texttt{TRECS} $n(z)$ irrespective of the choice of galaxy bias model. Major differences between the recovered and input distributions manifest primarily in the high-redshift tail, including features such as the bump around $z \sim 1.7$. A more accurate reconstruction of the input redshift distribution will require larger number of tomographic bins, which will be possible with surveys like \textit{Euclid} (\citealt{2025A&A...697A...1E}, see \citealt{2025arXiv251122732P} for such a pilot analysis), the Vera C. Rubin Observatory LSST \citep{2009arXiv0912.0201L, 2019ApJ...873..111I}, and 4MOST \citep{2019Msngr.175....3D}. These results from simulations provide a strong validation for our methodology and motivates recovering at least the peak of the true EMU-PS1 redshift distribution from real data.

\subsection{Effective incompleteness}\label{sec:incompleteness_test}

\begin{figure*}[!hbt]
    \begin{subfigure}{0.25\linewidth}
        \centering
        \includegraphics[width=\linewidth]{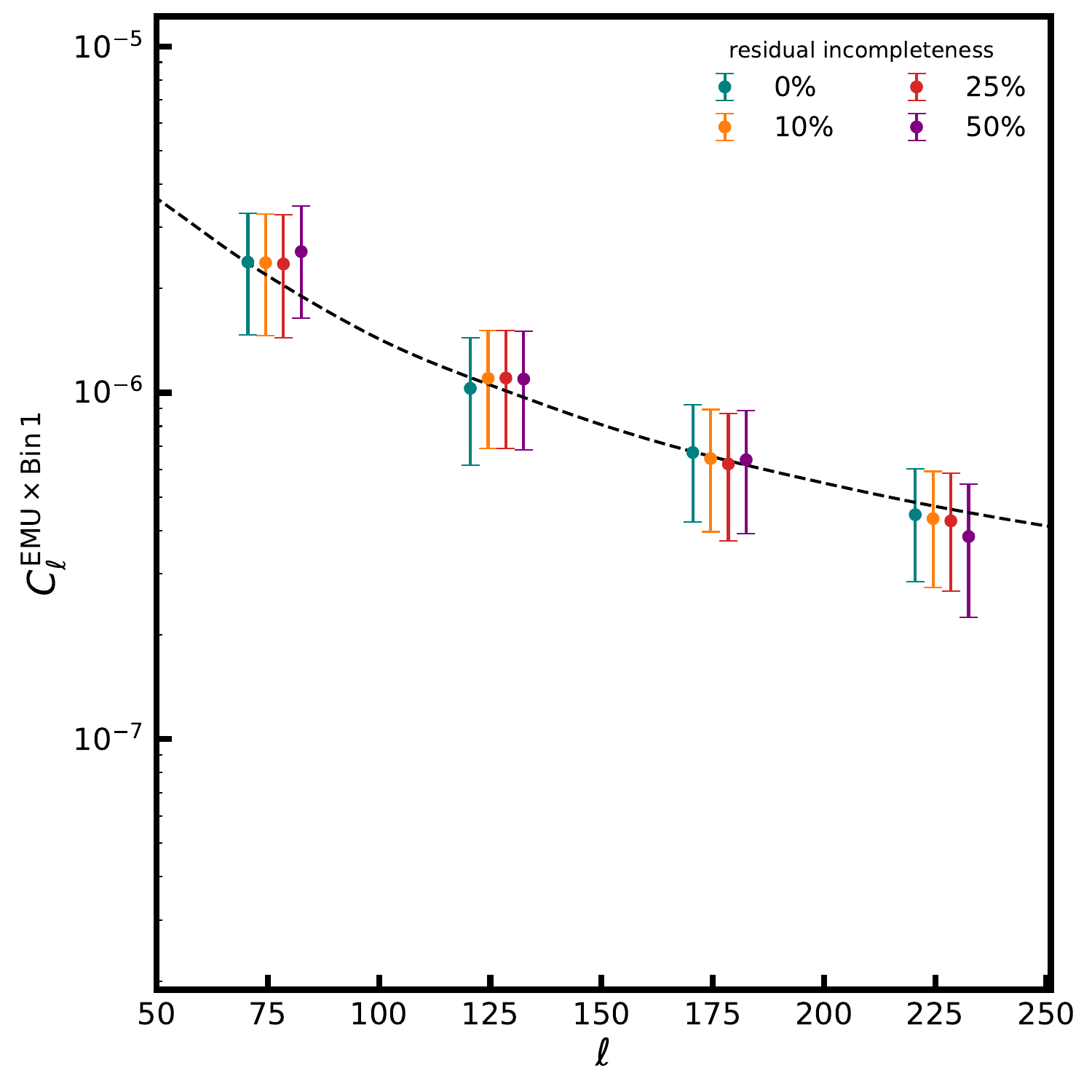}
    \end{subfigure}%
    \begin{subfigure}{0.25\linewidth}
        \centering
        \includegraphics[width=\linewidth]{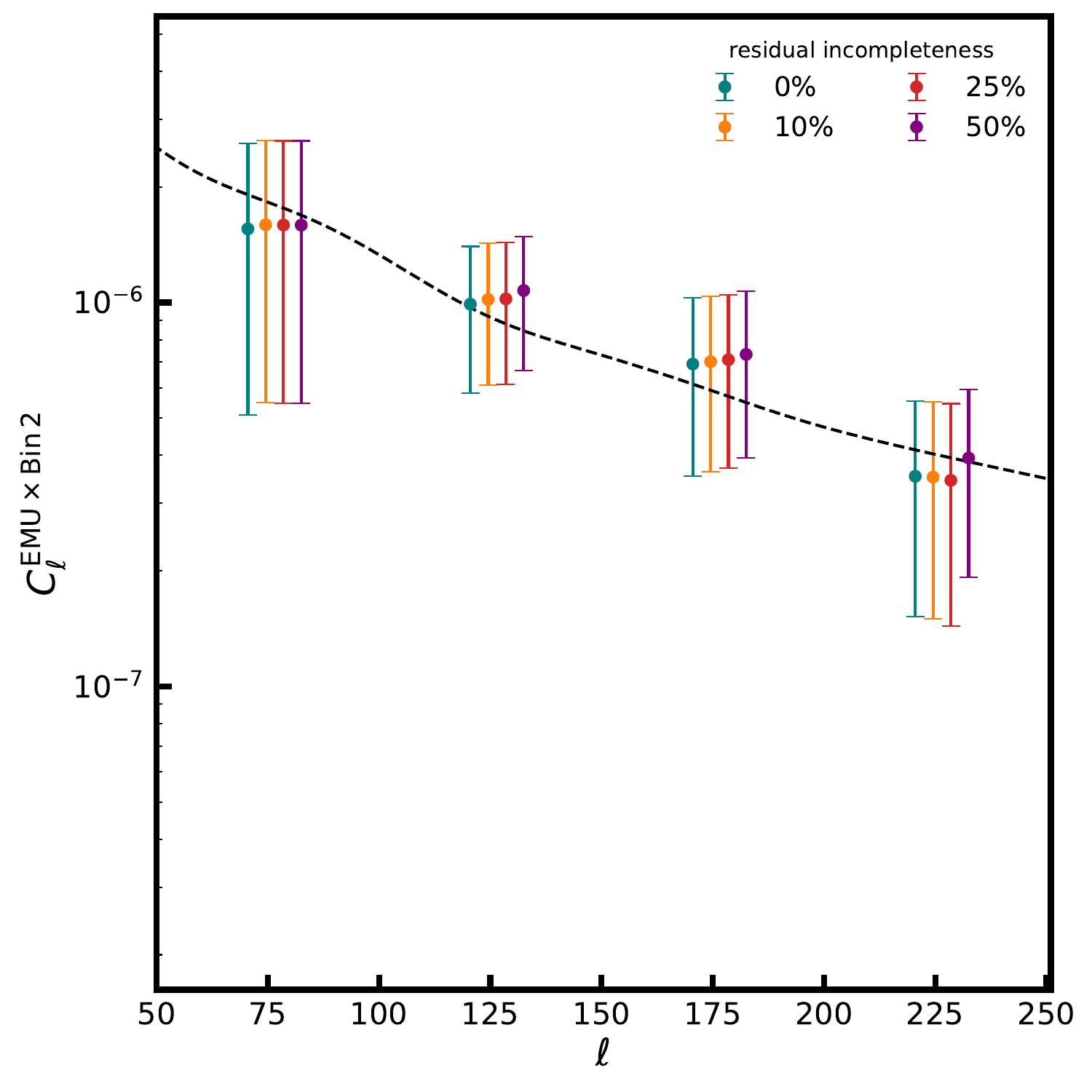}
    \end{subfigure}%
    \begin{subfigure}{0.25\linewidth}
        \centering
        \includegraphics[width=\linewidth]{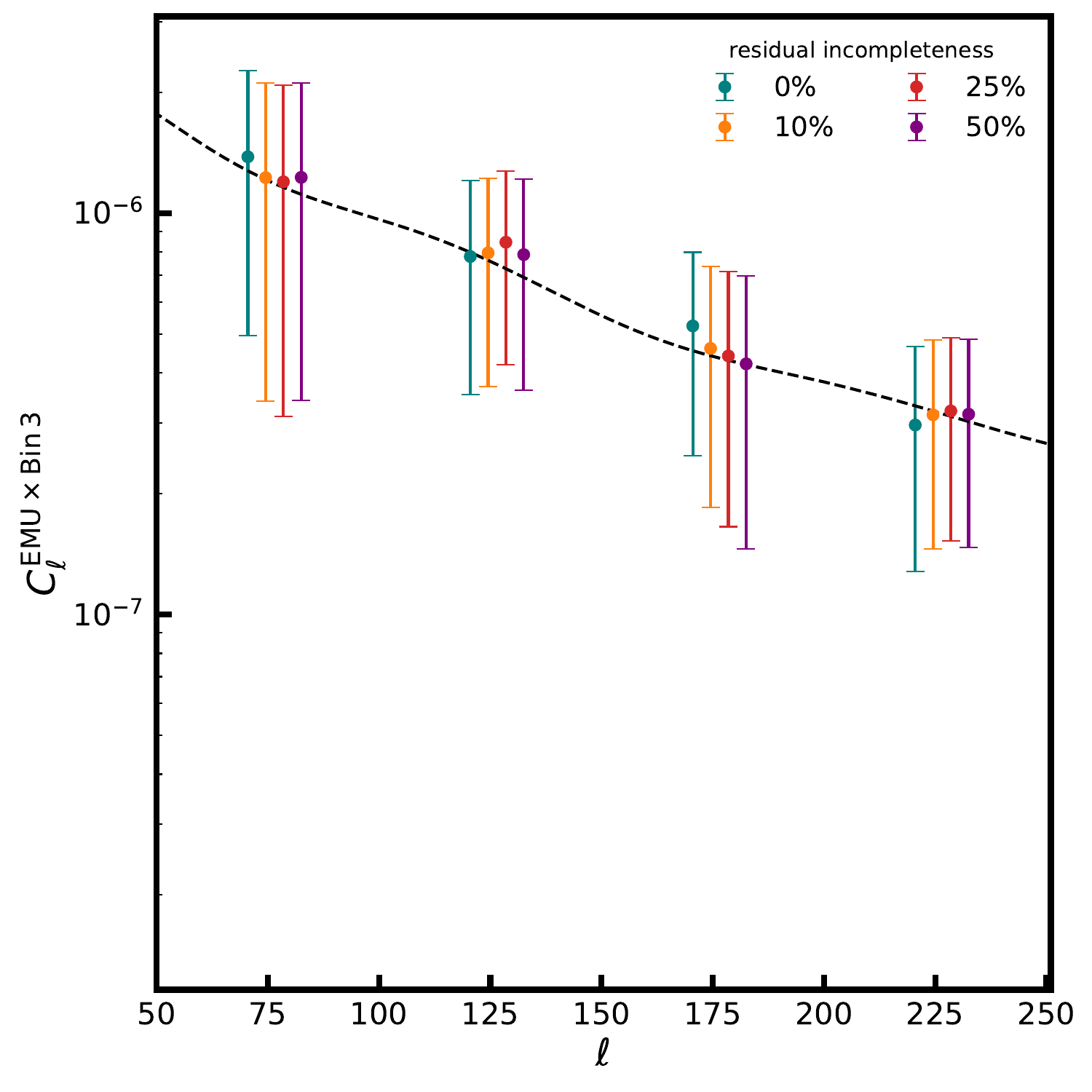}
    \end{subfigure}%
    \begin{subfigure}{0.25\linewidth}
        \centering
        \includegraphics[width=\linewidth]{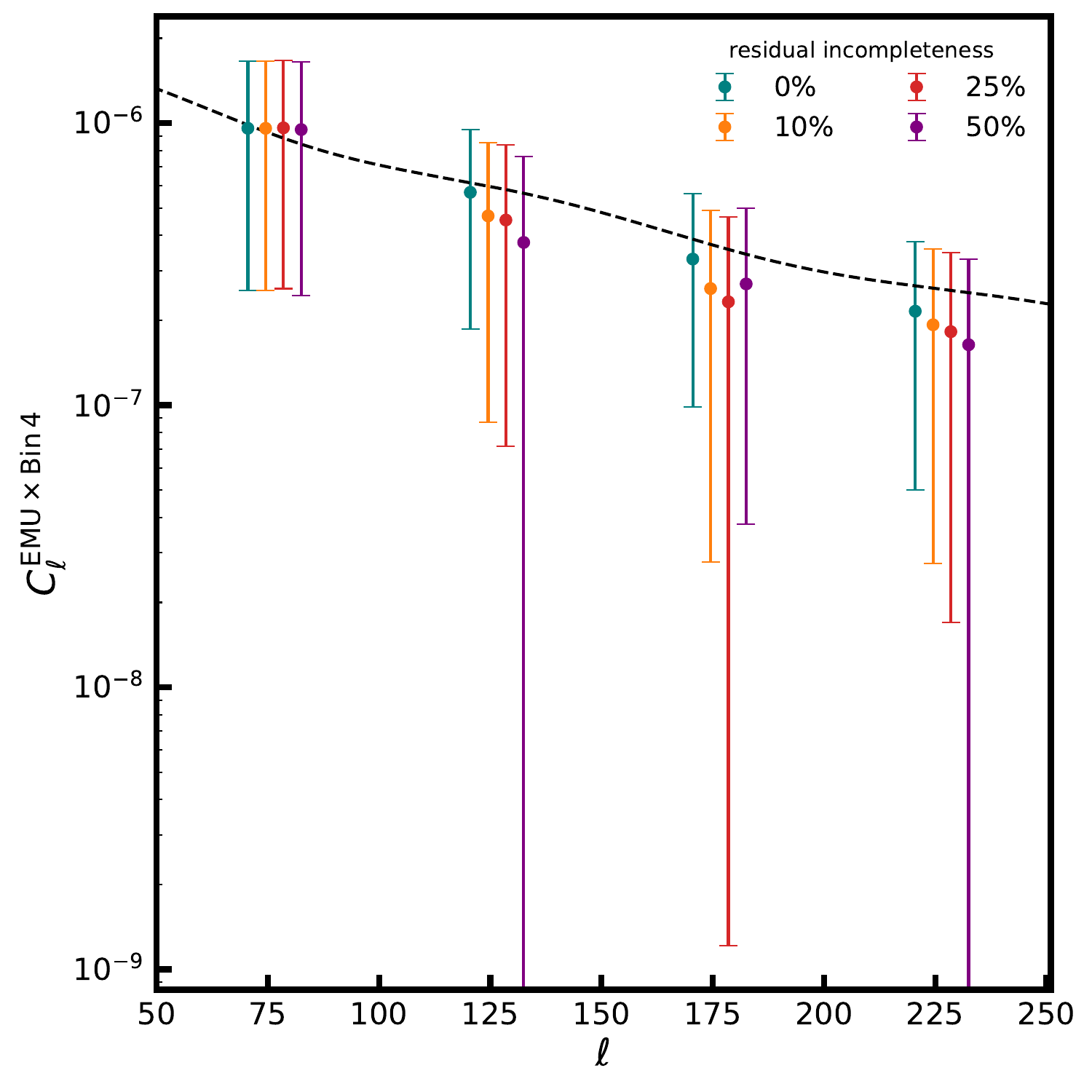}
    \end{subfigure}\\[5ex]
    \begin{subfigure}{0.25\linewidth}
        \centering
        \includegraphics[width=\linewidth]{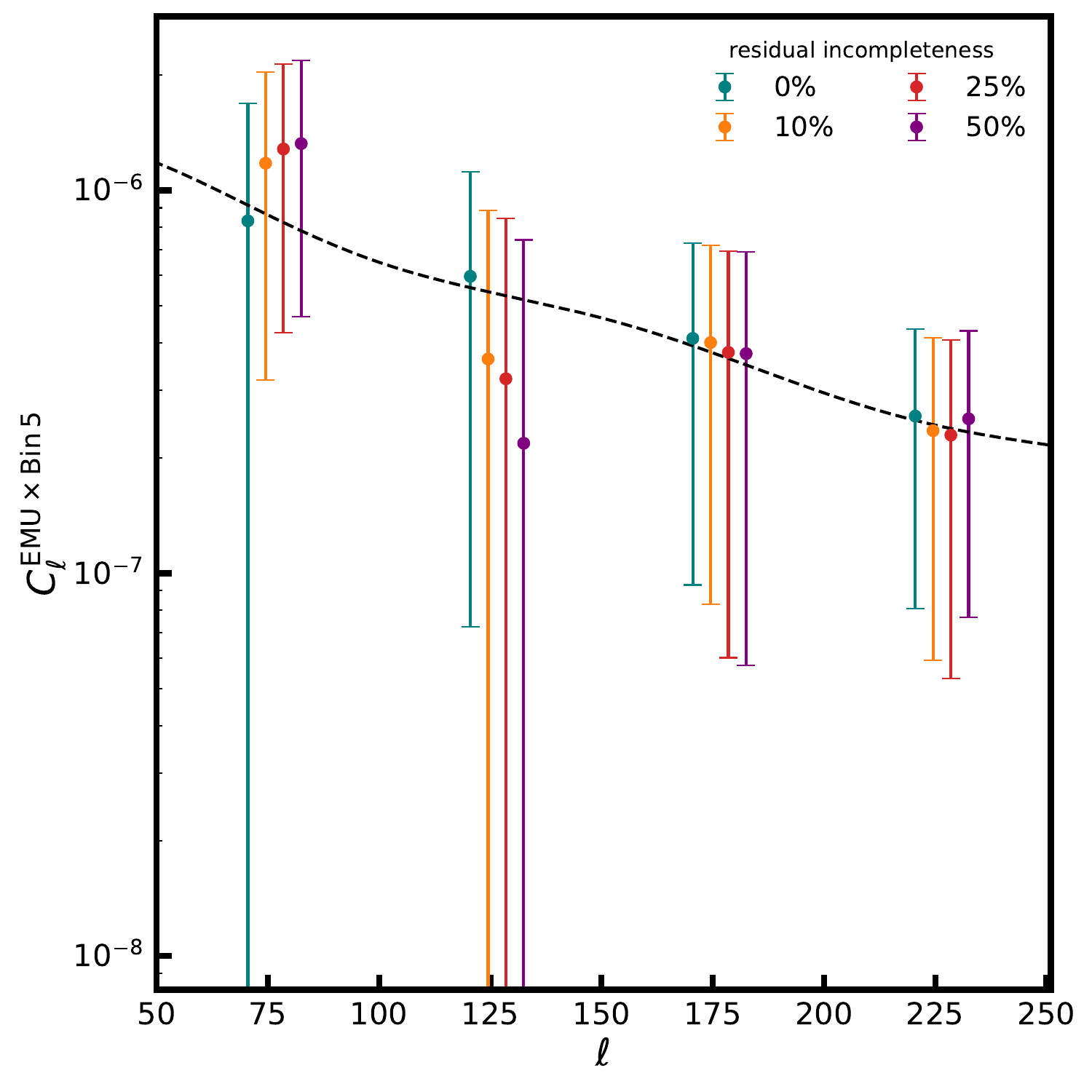}
    \end{subfigure}%
    \begin{subfigure}{0.25\linewidth}
        \centering
        \includegraphics[width=\linewidth]{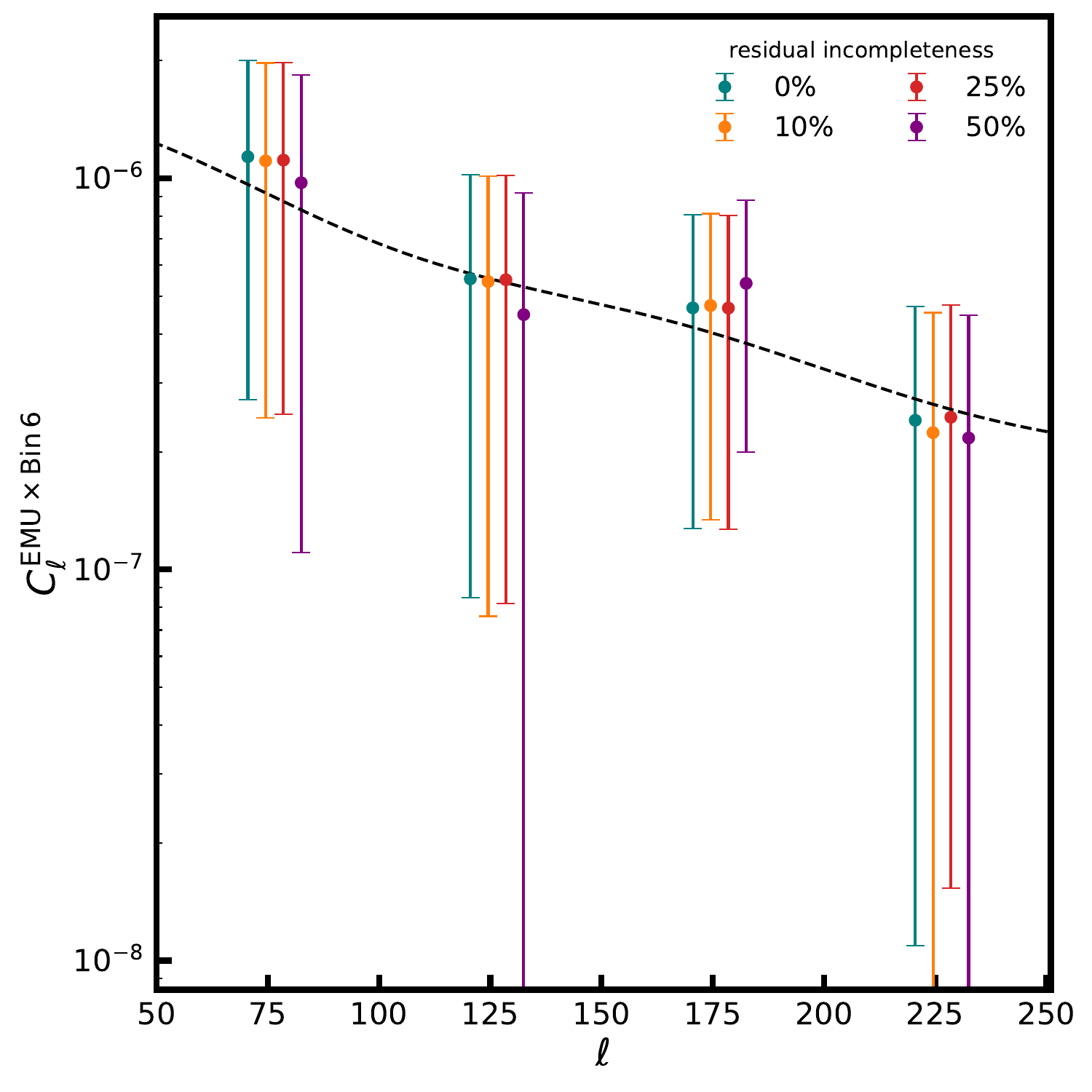}
    \end{subfigure}%
    \begin{subfigure}{0.25\linewidth}
        \centering
        \includegraphics[width=\linewidth]{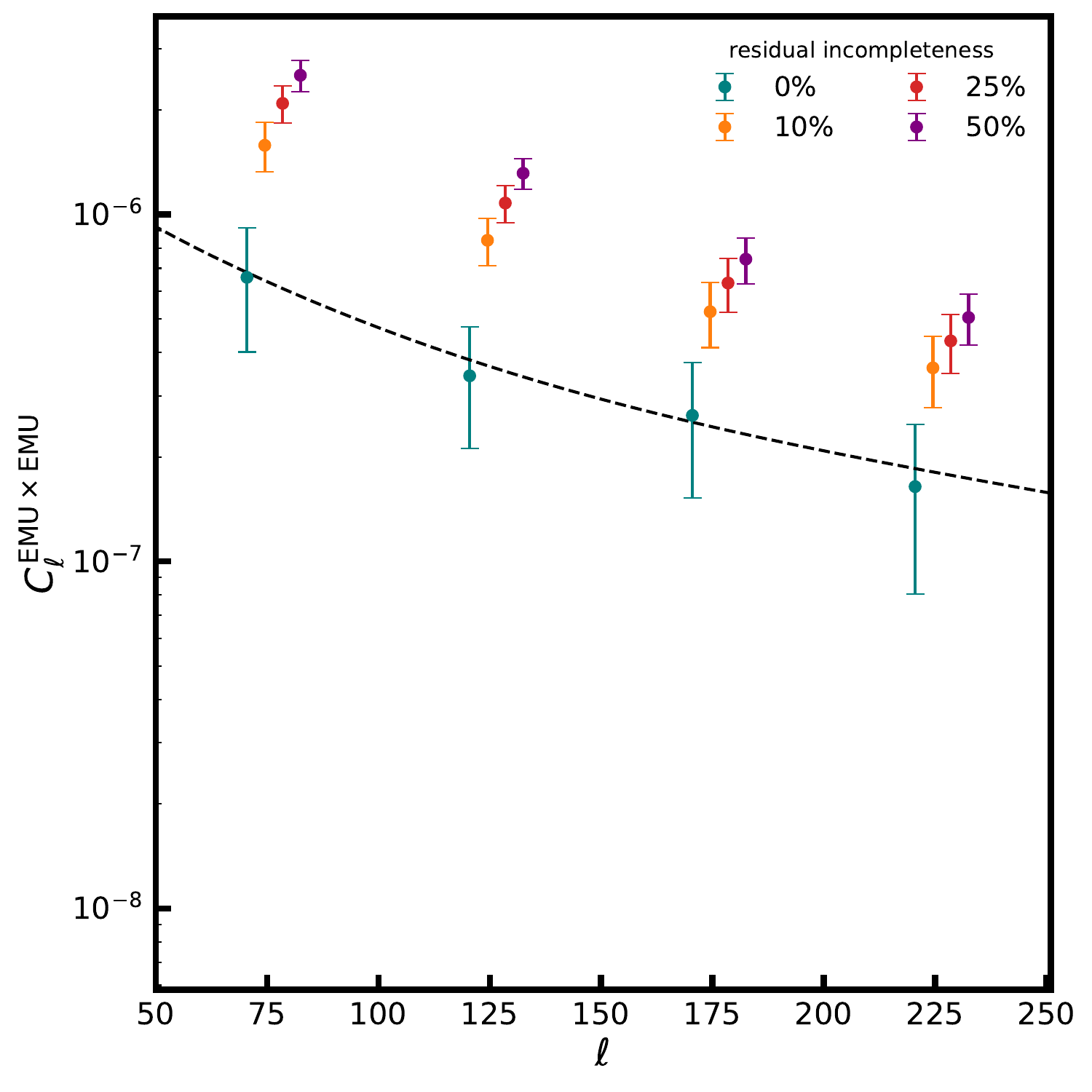}
    \end{subfigure}
    \caption{Average auto-power spectrum and cross-power spectra from 25 simulations of EMU-PS1 source catalogue and six DES tomographic bins, with different levels of residual incompleteness. The black dashed lines show the fiducial angular power spectra used for simulations.}
    \label{fig:incompleteness_validation}
\end{figure*}

An important systematic that can bias redshift distribution estimates derived through cross-correlation is the incompleteness of the radio source catalogue. Incompleteness arises from a combination of observational and instrumental effects, including flux detection thresholds that vary across the survey footprint, non-uniform noise levels, sidelobe contamination, confusion noise at faint flux densities, and incompleteness in source detection algorithms. \cite{2023PASA...40...28B, 2023PASA...40...27B} suggest that the EMU-PS1 sources are only $\sim 50-60\%$ complete at our flux threshold of $180\,\mu$Jy. We constructed weight maps for EMU-PS1 to account for incompleteness at $180$ and $400\,\mu$Jy flux cuts. In this section, we assess the impact of any residual incompleteness on the cross-power spectra between EMU-PS1 and DES tomographic bins.

We generated mock EMU-PS1 source catalogue and DES tomographic galaxy fields following the same framework outlined in Section \ref{sec:simulations}, with the additional step of assigning total intensity flux densities at 943 MHz to the simulated EMU galaxies by sampling from the flux distribution provided by \texttt{TRECS}. We used the PyBDSF generated EMU-PS1 RMS map to convert these fluxes to signal-to-noise ratios (SNR) for each simulated source, and apply a spatially dependent cut on the number of sources based on RMS map. We do not have peak flux densities for our simulated sources, but at 15" resolution of EMU-PS1 the majority of the faint sources will be compact. So as an approximation we preferentially remove sources at lower SNR based on a sigmoid function, where the probability of survival for a source is 50\% at SNR=5 and 100\% from SNR$>=10$. The RMS-based removal of sources is performed to quantify any scale dependent fluctuations in the cross-power spectra between EMU-PS1 and DES tomographic bins, that can influence our estimates of EMU-PS1 redshift distribution and galaxy bias.

We study four levels of residual incompleteness in the EMU data: 0\%, 10\%, 25\%, and 50\%. We created 25 mock catalogues of EMU-PS1 and the six DES tomographic bins. The average cross-power spectra and EMU-PS1 auto-power spectrum from these simulation are shown in Figure \ref{fig:incompleteness_validation}. We find that the cross-power spectra in all tomographic bins vary within $1\sigma$ uncertainties, but importantly do not produce any significant offsets due to different levels of residual incompleteness. The maximum impact is observed in the EMU-PS1 auto-power spectrum where residual incompleteness tends to increase the power on all scales considered in our analysis. However, since we employ only the cross-power spectra to estimate EMU-PS1 $n(z)$ and $b(z)$, we expect our results to be robust against any residual incompleteness not accounted for by the EMU-PS1 weight maps.

\section{Results}\label{sec:results}

\subsection{DES galaxy bias}\label{sec:des_gal_bias}
\begin{figure}[hbt!]
\centering
\includegraphics[width=\linewidth]{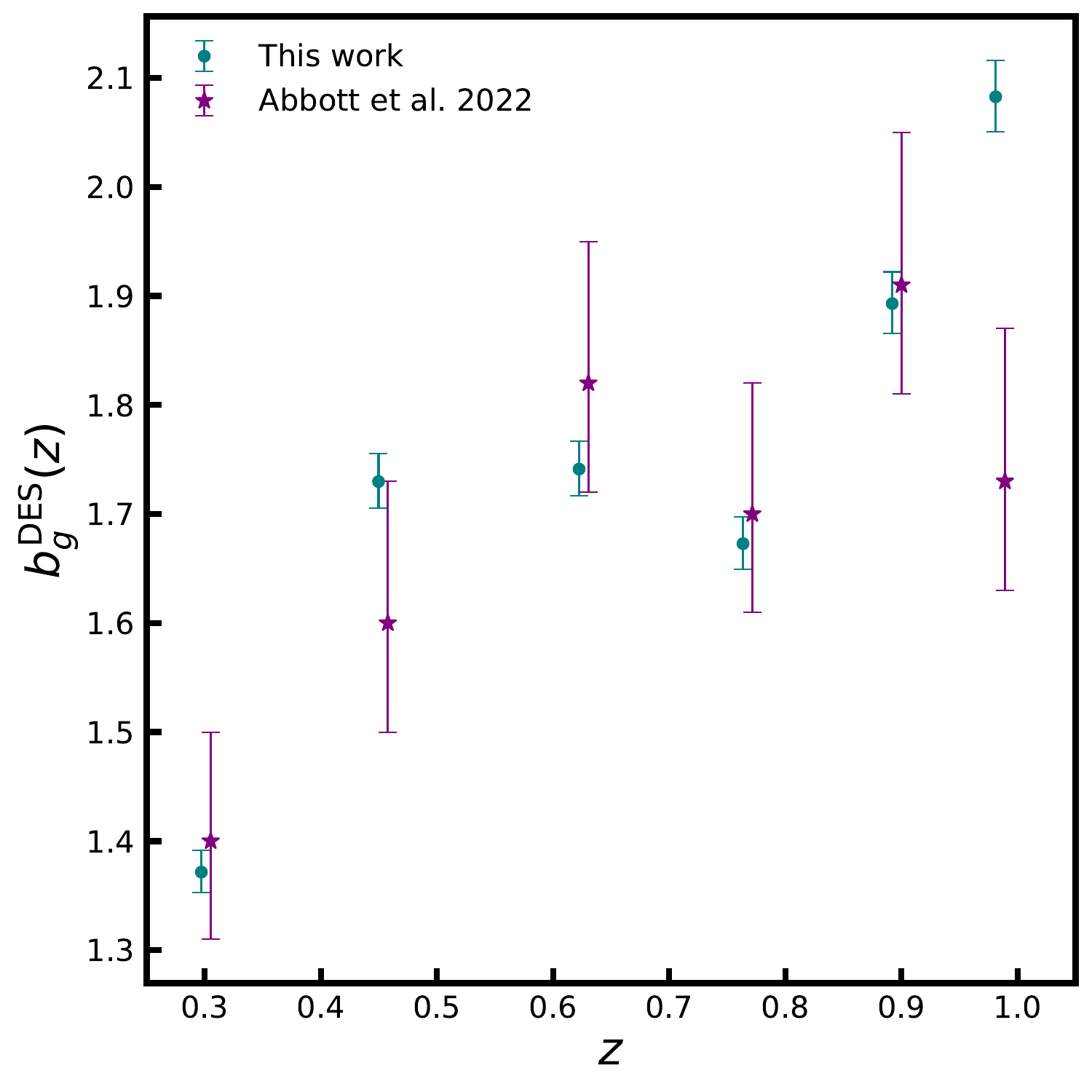}
\caption{Galaxy bias estimated for the six DES tomographic bins. The effective redshift is shifted for illustrative purposes only.}
\label{fig:galaxy_bias_des}
\end{figure}

Before estimating EMU-PS1 galaxy bias and redshift distribution, we need precise constraints on the DES galaxy bias. We start by binning the measured tomographic $C_{\ell}^{\text{DES}}$ power-spectra in eight linear multipole bins between $2\leq \ell\leq 250$. We then use maximum likelihood estimation framework (see section \ref{sec:parameter_estimation}) to determine linear galaxy bias in each DES tomographic bin. As shown in Figure \ref{fig:galaxy_bias_des}, our estimates of DES linear galaxy bias are consistent with previous measurements from \cite{2022PhRvD.105b3520A} within $1\sigma$ uncertainties, except for the last tomographic bin. We note that while \cite{2022PhRvD.105b3520A} employed a full $3\times 2$-point analysis to estimate galaxy bias, our results are based solely on galaxy clustering measurements. Also, the galaxy bias estimated from galaxy-galaxy lensing is found to be smaller than galaxy clustering measurements in the last MagLim tomographic bin as discussed in \cite{2022PhRvD.105b3520A, 2023PhRvD.107b3530C}. Table \ref{tab:total_chi_square_emu_des} lists the best-fit total chi-square value for each of the six DES tomographic bins as a measure of their goodness of fit.

For the purposes of this study, a linear galaxy bias model for the DES tomographic bins is adequate. However, future analyses using the EMU main survey will benefit from a larger sky area. To optimally utilize the cross-correlation measurements, a non-linear prescription for DES galaxy bias (as used in \citealt{2022PhRvD.106j3530P}) will be necessary to improve constraints on the redshift distribution of radio sources.

\begin{figure*}[thb!]
    \begin{subfigure}[b]{\linewidth}
        \includegraphics[width=\linewidth]{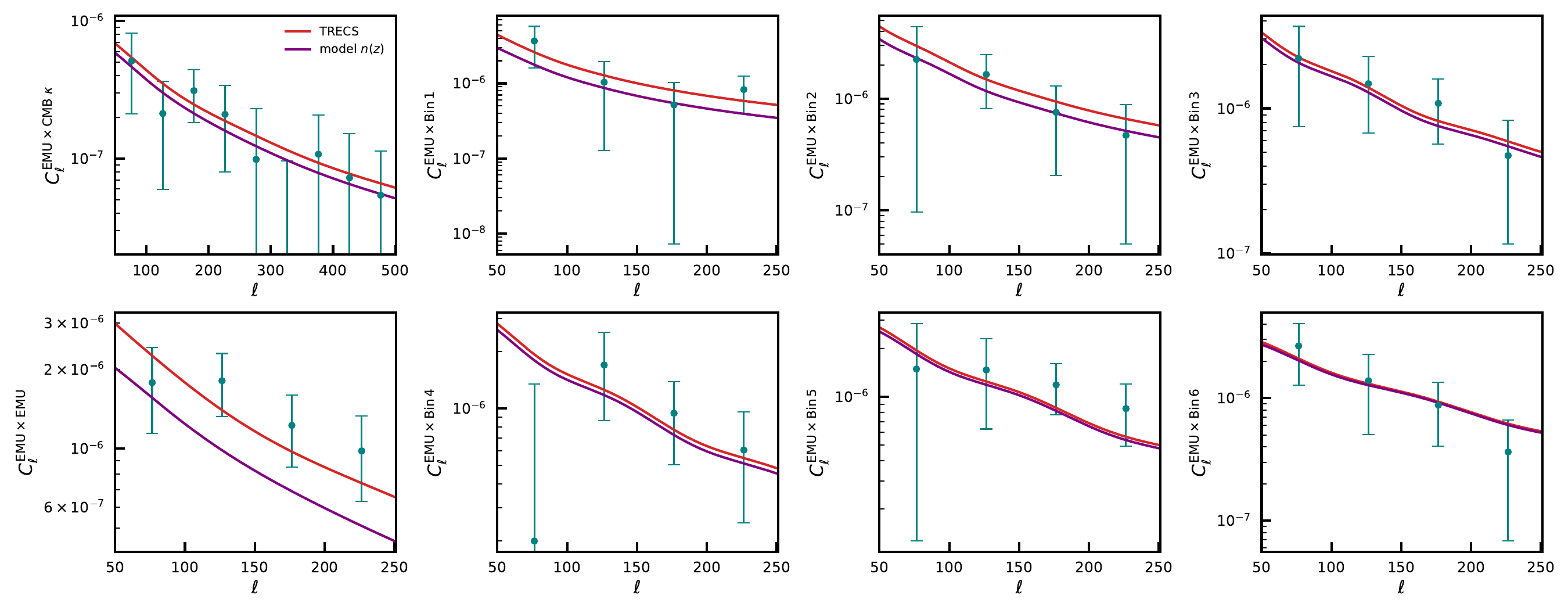}
        \caption{$180\,\mu$Jy}
        \label{fig:power_spectra_emu_pybdsf_cross_des_180}
    \end{subfigure}\\[5ex]
    \begin{subfigure}[b]{\linewidth}
        \includegraphics[width=\linewidth]{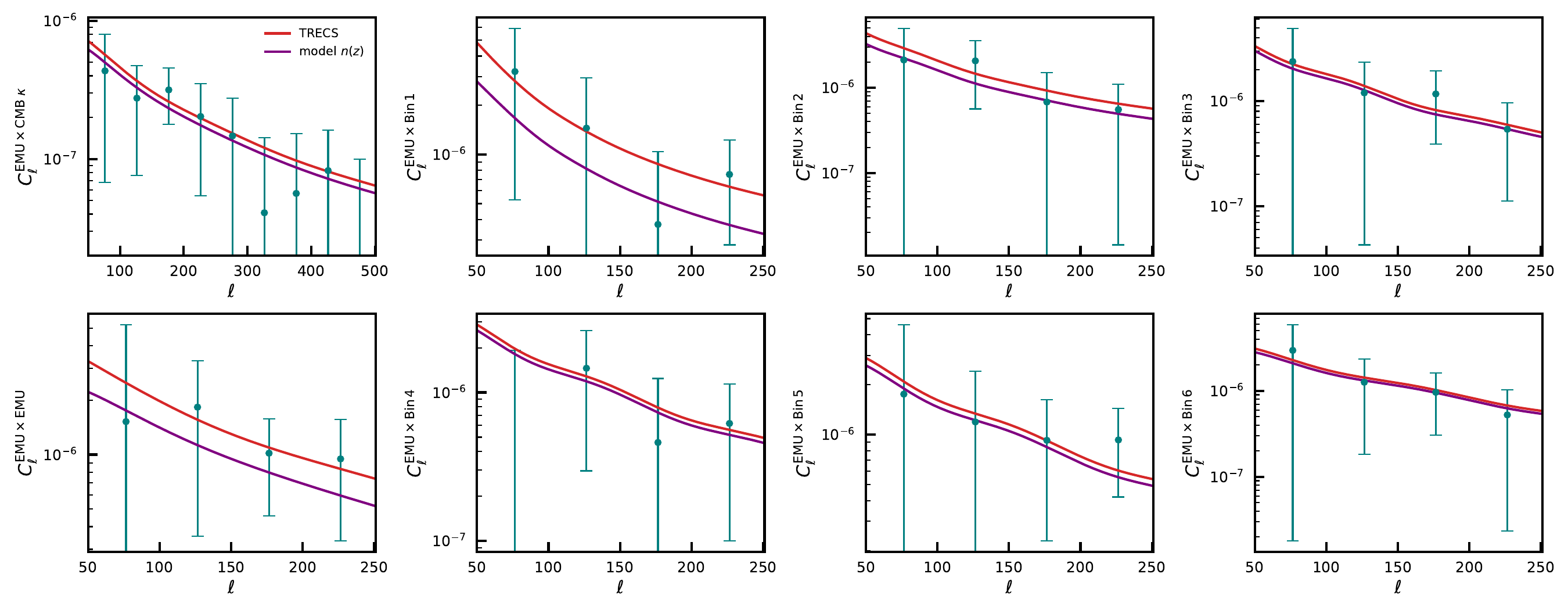}
        \caption{$400\,\mu$Jy}
        \label{fig:power_spectra_emu_pybdsf_cross_des_400}
    \end{subfigure}
    \caption{The angular auto power spectra and cross-power spectra measured using EMU-PS1, \textit{Planck} CMB lensing convergence and DES tomographic bins measured at (a) $180\,\mu$Jy and (b) $400\,\mu$Jy flux cuts. The best-fit power spectra obtained using \texttt{TRECS} $n(z)$ and mixture model for galaxy bias is shown with red solid lines. The best-fits using LoTSS DR2 model $n(z)$ (Eq.\ \ref{eq:model_nz_lotssdr2}) are shown with the purple curves. The $1\sigma$ errors on data points are computed from the diagonal of the sample covariance matrix (Eq. \ref{eq:sample_covariance}). We see the failure of both the $n(z)$ model and \texttt{TRECS} to fit the cross-power spectrum on large-scales in bin 4, leading to the large $\chi^2$ value given in table \ref{tab:total_chi_square_emu_des}.}
    \label{fig:power_spectra_emu_pybdsf_cross_des}
\end{figure*}

\subsection{Power spectra and detection significance}

The EMU-PS1 auto-power spectrum ($C_{\ell}^{\text{EMU}}$), cross-spectra with \textit{Planck} CMB lensing convergence ($C_{\ell}^{\text{EMU}\times\kappa}$) and the cross-power spectra with DES tomographic bins ($C_{\ell}^{\text{EMU}\times\text{DES}}$) are shown in Figure~\ref{fig:power_spectra_emu_pybdsf_cross_des_180} for $180\,\mu$Jy flux cut and in Figure~\ref{fig:power_spectra_emu_pybdsf_cross_des_400} for $400\,\mu$Jy flux cut. $C_{\ell}^{\text{EMU}\times\kappa}$ is binned into nine linear multipole bins between $50\leq\ell\leq500$. The other power spectra are binned into four linear multipole intervals with bin width between $50\leq\ell\leq250$. We apply a conservative upper multipole cut at $\ell=250$ in $C_{\ell}^{\text{EMU}}$ and $C_{\ell}^{\text{EMU}\times\text{DES}}$ to restrict our analysis to linear scales. Beyond this range, the contribution from multi-component radio sources also begins to dominate the EMU-PS1 power spectrum (see T25), and a detailed treatment to mitigate these effects is beyond the scope of this work.

The standard $1\sigma$ uncertainties on the power spectra measurements are computed from the diagonal of the sample covariance matrix defined in Eq. \ref{eq:sample_covariance}. Assuming the null hypothesis of zero correlation, the significance of detecting the $C_{\ell}^{\text{EMU}\times\text{DES}}$ power spectra was computed using $\smash{\sqrt{\chi^2_{\text{null}}-\chi^2_{\text{best-fit}}}}$. The resulting detection significances for the cross-power spectra with six DES tomographic bins are $3.8\sigma,\,4.6\sigma,\,3.8\sigma,\,3.5\sigma,\,5.5\sigma$ and $3.9\sigma$, respectively.

\subsection{EMU-PS1 redshift distribution and galaxy bias}

\begin{figure*}[!hbt]
    \begin{subfigure}{0.5\linewidth}
        \centering
        \includegraphics[width=\linewidth]{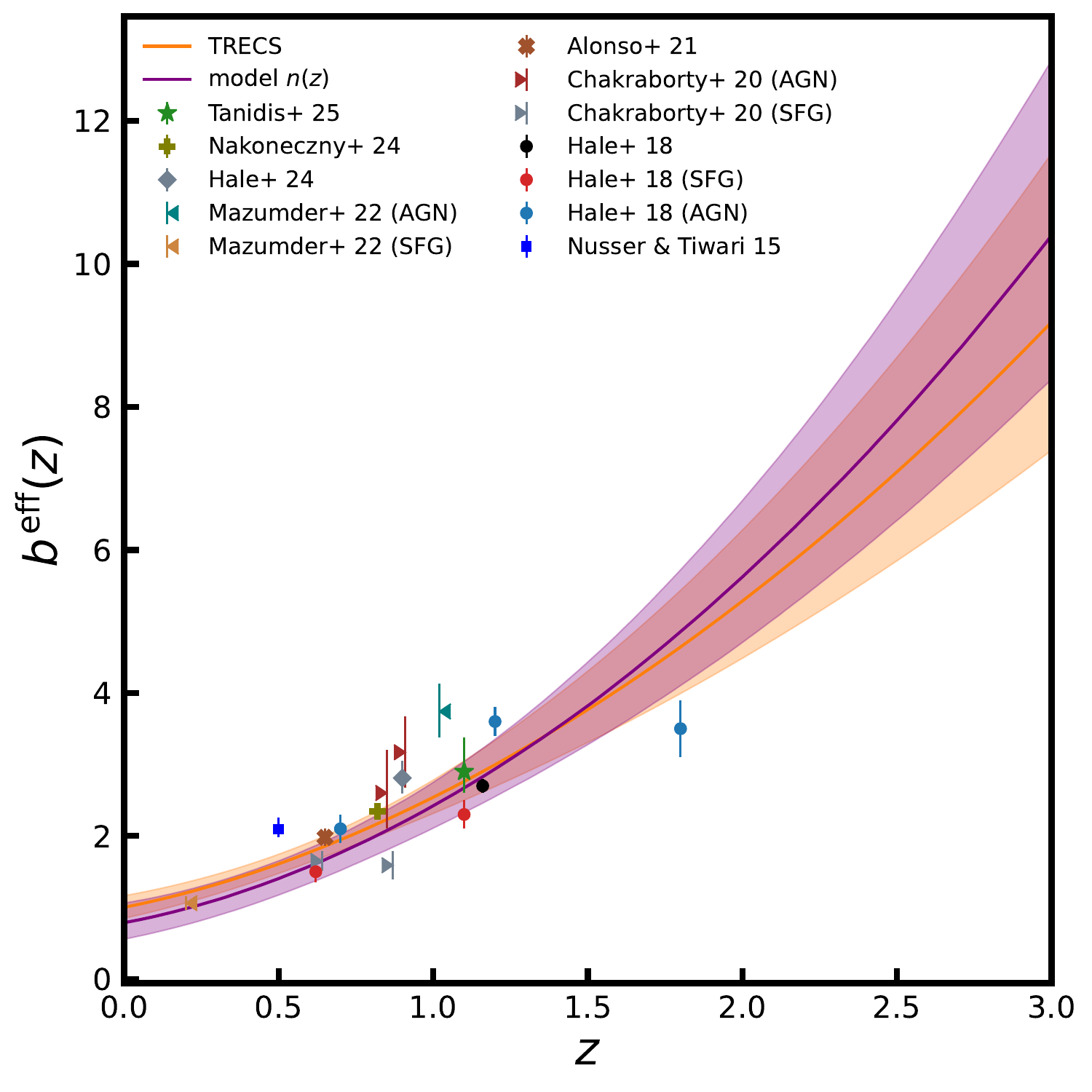}
        \caption{$180\,\mu$Jy}
    \end{subfigure}%
    \begin{subfigure}{0.5\linewidth}
        \centering
        \includegraphics[width=\linewidth]{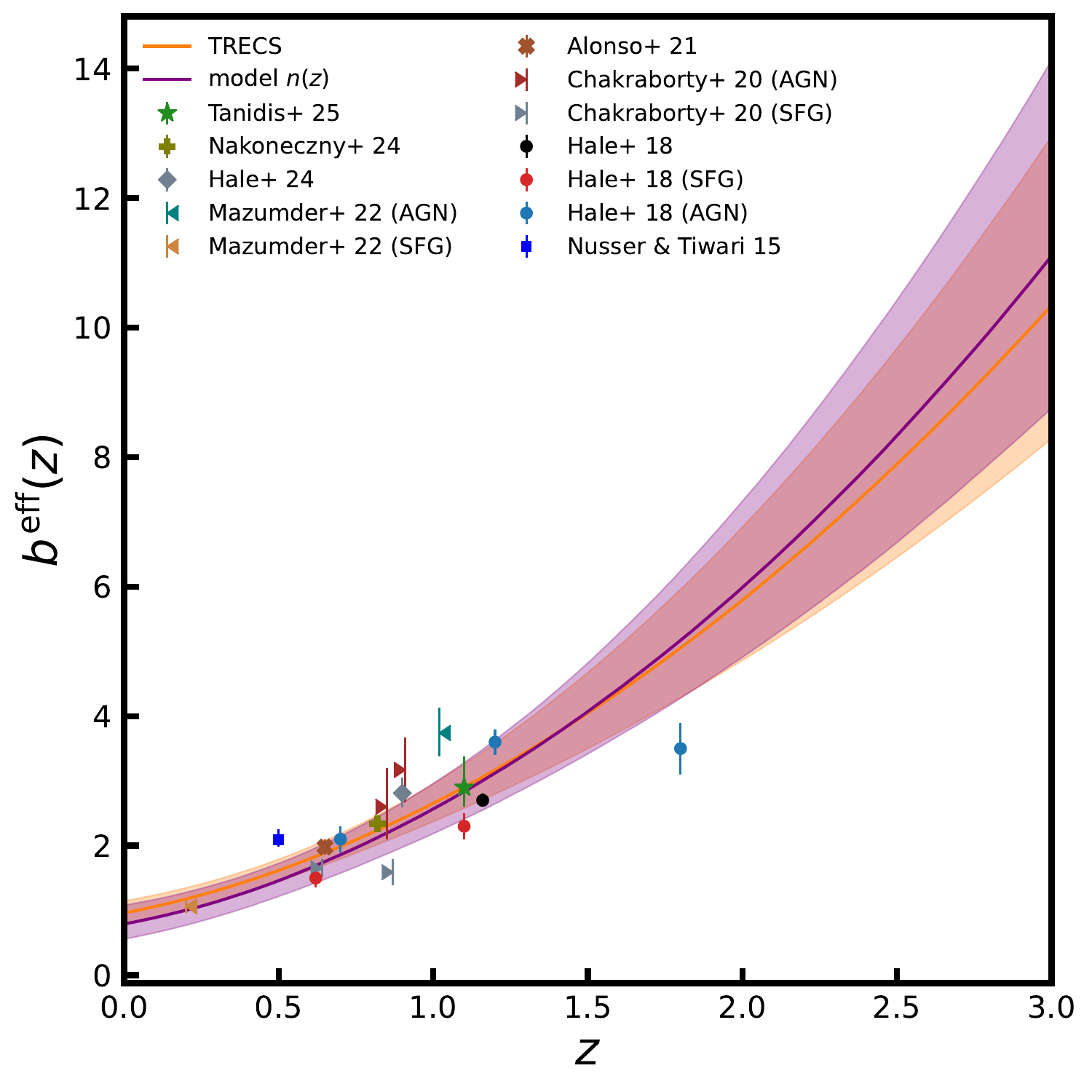}
        \caption{$400\,\mu$Jy}
    \end{subfigure}
    \caption{Best-fit galaxy bias (in solid lines) with $\pm 1\sigma$ uncertainty region with \texttt{TRECS} (in orange) and model $n(z)$ (in purple) for (a) $180\,\mu$Jy and (b) $400\,\mu$Jy flux cuts. The markers show the galaxy bias estimates for radio galaxies available in the literature \citep{2015ApJ...812...85N, 2018MNRAS.474.4133H, 2020MNRAS.494.3392C, 2021MNRAS.502..876A, 2022MNRAS.517.3407M, 2024MNRAS.527.6540H, 2024A&A...681A.105N, 2025PASA...42...62T}.}
    \label{fig:best_fit_gal_bias}
\end{figure*}

\begin{figure*}[htb!]
    \begin{subfigure}[b]{0.5\linewidth}
        \includegraphics[width=\linewidth]{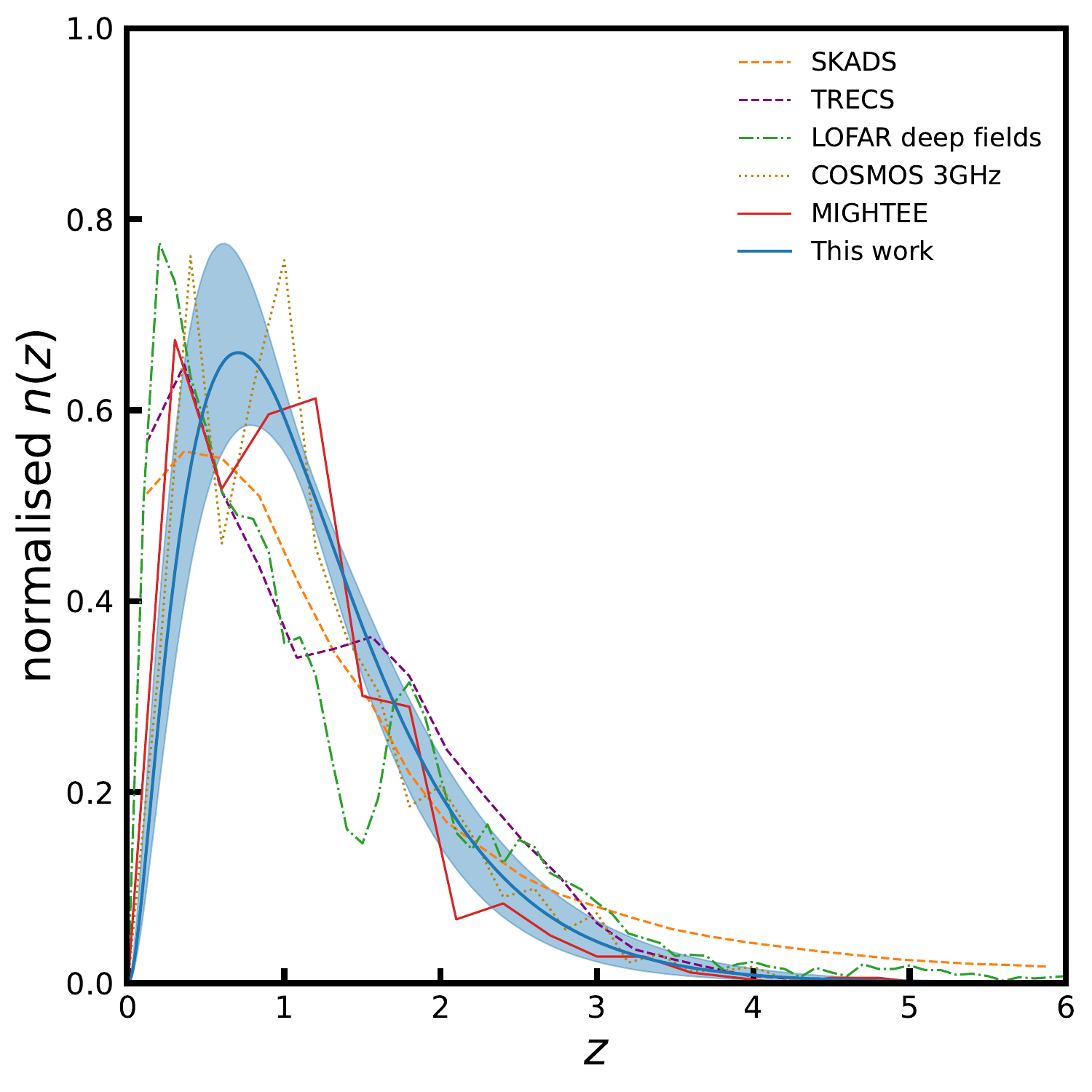}
        \caption{$180\,\mu$Jy}
    \end{subfigure}%
    \begin{subfigure}[b]{0.5\linewidth}
        \includegraphics[width=\linewidth]{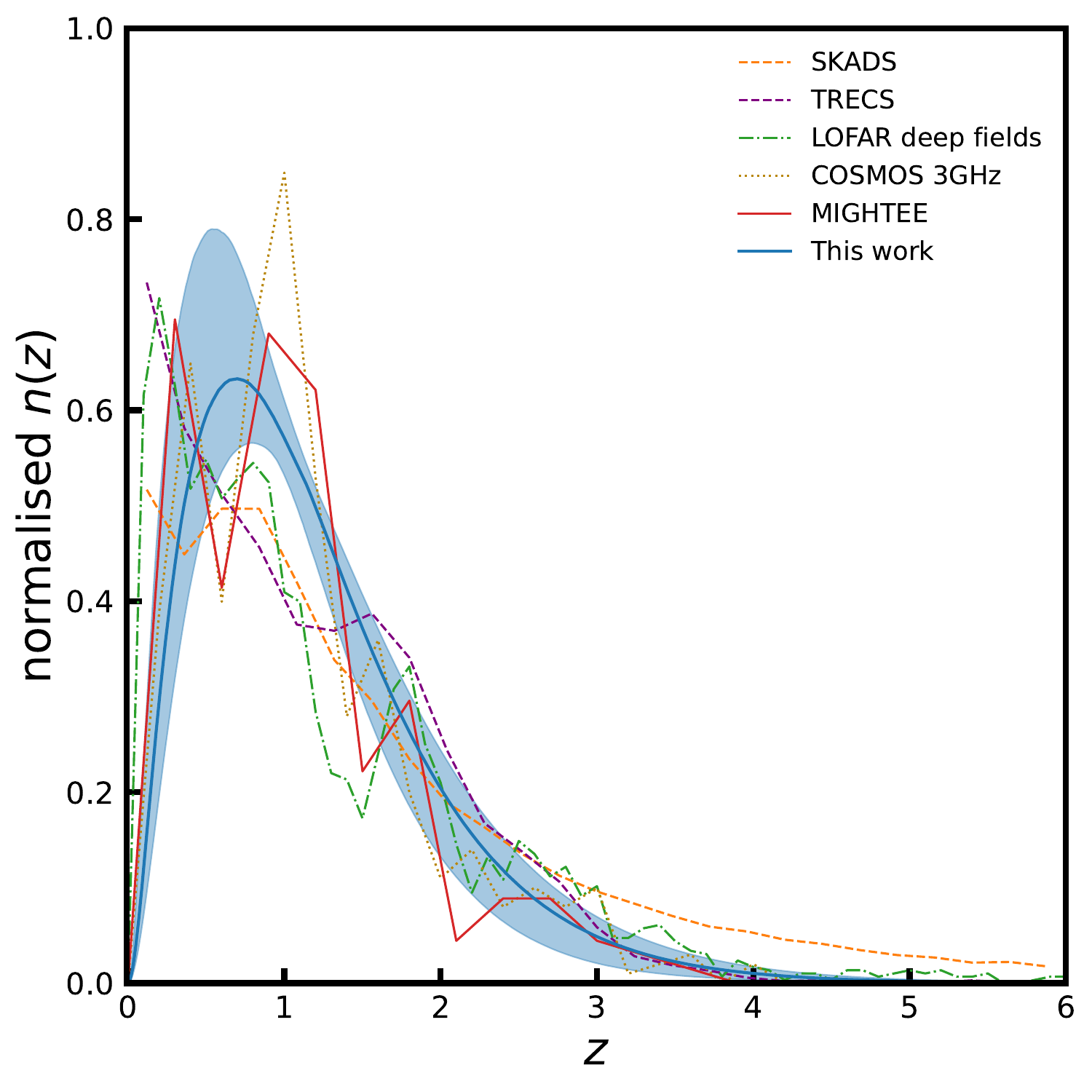}
        \caption{$400\,\mu$Jy}
    \end{subfigure}
    \caption{Best-fit redshift distribution assuming LoTSS DR2 model $n(z)$ (Eq. \ref{eq:model_nz_lotssdr2}) obtained from cross-correlating EMU-PS1 sources at (a) $180\,\mu$Jy and (b) $400\,\mu$Jy flux cuts with DES MagLim galaxy sample (shown with solid blue curve) in linear (\textit{left}) and logarithmic (\textit{right}) x-axis. The \texttt{TRECS} and \texttt{SKADS} $n(z)$ are shown in purple and orange, respectively. The redshift distributions from LOFAR deep fields (in green), COSMOS 3GHz (in yellow) and MIGHTEE survey (in red) are shown for comparison.}
    \label{fig:compare_nz_skads_trecs_model_nz}
\end{figure*}

\subsubsection{With a simulated $n(z)$ from \texttt{TRECS}}\label{sec:results_trecs}

We apply flux cuts at $180\,\mu$Jy and $400\,\mu$Jy in the \texttt{TRECS} simulated source catalogues to construct their appropriate redshift distributions and compute the ratio of SFGs to AGNs in the six DES redshift bins. We then estimate $b_{\text{SFG},0}$ and $b_{\text{AGN},i}$ using $C_{\ell}^{\text{EMU}\times\text{DES}}$ power spectra only. We then use $C_{\ell}^{\text{EMU}}$ and $C_{\ell}^{\text{EMU}\times\kappa}$ to validate the resulting best-fit parameters. The best results are obtained with a quadratic AGN galaxy bias model for both $180\,\mu$Jy and $400\,\mu$Jy flux cuts.

The best-fit theoretical power spectra using \texttt{TRECS} $n(z)$ are shown in Figure \ref{fig:power_spectra_emu_pybdsf_cross_des}. We find that the cross-power spectra $C_{\ell}^{\text{EMU}\times\text{DES}}$, $C_{\ell}^{\text{EMU}\times\kappa}$ and auto-power spectra $C_{\ell}^{\text{EMU}}$ are fit very-well by \texttt{TRECS} $n(z)$. We obtained a total $\chi^{2}$ value of $7.78$ for $180\,\mu$Jy flux cut and value of $3.15$ for $400\,\mu$Jy, both with $\nu=20$ degrees of freedom. It is important to note that we relied on the ratio of AGNs to SFGs in the six tomographic bins to fit the cross-power spectra $C_{\ell}^{\text{EMU}\times\text{DES}}$ with \texttt{TRECS} $n(z)$. The best-fit $\chi^{2}$ values from individual fits, using the \texttt{TRECS} prediction, are listed in Table~\ref{tab:total_chi_square_emu_des}. In Figure \ref{fig:best_fit_gal_bias}, we show the best-fit effective galaxy bias for \texttt{TRECS} with $\pm 1\sigma$ uncertainty.

\subsubsection{With model \texorpdfstring{$n(z)$}{Lg}}\label{sec:results_model}

We adopted the model redshift distribution from LoTSS DR2 analysis \citep{2024A&A...681A.105N} with parametric form given in Eq. \ref{eq:model_nz_lotssdr2}, to infer the distribution of EMU-PS1 radio sources. In Figure \ref{fig:compare_nz_skads_trecs_model_nz}, we compare our resulting best-fit redshift distribution (in blue) with other radio continuum surveys after applying appropriate flux thresholds (assuming a spectral index of 0.75) for $180\,\mu$Jy and $400\,\mu$Jy flux cuts. The shaded region shows the $1\sigma$ uncertainty in our best-fit EMU-PS1 $n(z)$. Notably, most of the variation lies in the amplitude rather than the shape of the distribution. Our best-fit EMU-PS1 $n(z)$ peaks at $z\sim 0.85$. which is higher than those obtained from \texttt{TRECS} and LOFAR deep fields\citep{2021A&A...648A...4D}, but is consistent with surveys like COSMOS 3GHz \citep{2017A&A...602A...6S} and the MeerKAT International GHz Tiered Extragalactic Exploration (MIGHTEE) Survey \citep{2024MNRAS.527.3231W}. The redshifts of MIGHTEE sources are based on visual cross-match with the UltraVISTA $K_{s}$ band selected catalogue \citep{2012A&A...544A.156M}. Their analysis reveals a significantly larger number of AGNs at $z \sim 1$ than predicted by simulations, shifting the distribution’s peak to $z \sim 0.94$.

Figure \ref{fig:best_fit_gal_bias} shows the best-fit effective bias for our model $n(z)$ with $1\sigma$ uncertainty. The best results adopting our model $n(z)$ are obtained with the quadratic polynomial galaxy bias model at both $180\,\mu$Jy and $400\,\mu$Jy flux cuts. The galaxy bias evolution obtained with our model $n(z)$ is very consistent with that from \texttt{TRECS}, even without using any information about the relative contribution of AGNs to SFGs. Our estimates of EMU-PS1 galaxy bias is also consistent with galaxy bias estimates for radio galaxies available in the literature \citep{2015ApJ...812...85N, 2018MNRAS.474.4133H, 2020MNRAS.494.3392C, 2021MNRAS.502..876A, 2022MNRAS.517.3407M, 2024MNRAS.527.6540H, 2024A&A...681A.105N, 2025PASA...42...62T}. The fits to the angular power spectra $C_{\ell}^{\text{EMU}\times\text{DES}}$, $C_{\ell}^{\text{EMU}\times\kappa}$ are comparable to that obtained using \texttt{TRECS}, as shown in Figure \ref{fig:power_spectra_emu_pybdsf_cross_des}. However, the fit to EMU-PS1 auto-power spectrum is poor in case of $180\,\mu$Jy flux cut, where our best-fit model yield smaller amplitudes than necessary to match the observed data. This indicates that there may be some residual systematics in the EMU-PS1 field that show up in the EMU-PS1 auto-power spectrum (for the $180\,\mu$Jy data), as demonstrated in Section \ref{sec:incompleteness_test}. For the joint-data vector of cross-power spectra $C_{\ell}^{\text{EMU}\times\text{DES}}$ with our model $n(z)$, we obtained a total $\chi^{2}$ of 9.25 at $180\,\mu$Jy and $\chi^{2}=4.03$ at $400\,\mu$Jy flux cut with $\nu=18$ degrees of freedom. The best-fit $\chi^{2}$ values from individual fits are listed in Table~\ref{tab:total_chi_square_emu_des}. We also present the chi-square values of the joint cross-power spectra data vector with different galaxy bias models in Table \ref{tab:total_chi_square_emu_des_diff_bias_models}.


In Section~\ref{sec:simulations}, we demonstrated that our methodology reliably recovers the true location of the peak of the EMU-PS1 $n(z)$ and that this result is robust against the galaxy bias model used in the cross-correlation analysis. We also investigated the effects of incompleteness via simulations and showed that we recover the location of the peak of the EMU-PS1 $n(z)$ across varying levels of completeness. This strengthens the credibility of our finding against internal systematics in EMU-PS1 data. We list below several plausible explanations for the differences in the inferred EMU-PS1 radio sources distribution using model $n(z)$:
\begin{itemize}
    \item Residual incompleteness in EMU-PS1: In Section \ref{sec:incompleteness_test}, we demonstrated that at $180\,\mu$Jy the simulated cross-power spectra $C_{\ell}^{\text{EMU}\times\text{DES}}$ are insensitive, whereas the auto-power spectra $C_{\ell}^{\text{EMU}}$ deviates significantly from its fiducial value under the influence of residual incompleteness in EMU-PS1 field. We also see a similar trend in measured angular power spectra (Figure \ref{fig:power_spectra_emu_pybdsf_cross_des_180}) at $180\,\mu$Jy where our best-fit model $n(z)$ and $b(z)$ fail to fit the EMU-PS1 auto-power spectra. However, since we adopted a RMS based removal of sources using SNR as a metric, our simulations may not fully capture the effects of residual incompleteness in the cross-power spectra resulting in high redshift peaks of our inferred EMU-PS1 redshift distributions.
    \item A greater population of radio-quiet AGNs at intermediate redshifts ($z \sim 1$–$2$): As suggested by MIGHTEE survey, the abundance of such AGNs at $z\sim 1-2$ may be underestimated in current simulations or overlooked in other radio continuum surveys. Because our method relies solely on measured cross-correlations between EMU-PS1 and DES galaxies, it is potentially more sensitive to the true redshift distribution of AGNs than those recovered from traditional cross-matching techniques.
    \item Missing cross-correlation data at $z<0.2$: Cross correlation data at low redshifts $z<0.2$ is crucial to put tight constraints on departures from redshift distributions from \texttt{TRECS} or \texttt{SKADS}. In absence of DES redshift bin between $0\leq z\leq 0.2$, there is room for biases in modelling the rising part of the radio redshift distribution using model $n(z)$, especially at $180\,\mu$Jy.
    \item Residual systematics in DES redshift distributions or galaxy bias evolution: Although we use the official DES $n(z)$ distributions, known systematics at $z > 0.85$ have been reported in the DES tomographic bins \citep{2022PhRvD.105b3520A}. Furthermore, uncertainties in photometric redshifts can bias the inferred mean and width of the distributions, leaving room for unidentified systematic errors that could affect our analysis.
    \item Redshift bin misclassification in DES due to photometric uncertainties: Errors in assigning galaxies to tomographic bins due to photometric redshift uncertainties can bias the inferred clustering signals and distort conclusions drawn from angular power spectra \citep{2024A&A...687A.150S, 2024A&A...690A.338S}.
    \item Survey area limitations and selection effects: Although less likely, it is possible that the relatively small area covered by EMU-PS1 ($\sim 270 \text{ deg}^{2}$) can result in selection effects and sample variance. Furthermore, the small area of EMU-PS1 reduces the number of data points for the measured cross-power spectra, resulting in uncertainties on inferred redshift distribution and galaxy bias evolution. 
\end{itemize}

A detailed investigation of these possibilities is beyond the scope of the present work. However, future analyses using the main EMU survey, with its significantly larger sky coverage, will be instrumental in testing these hypotheses and further constraining the redshift distribution of radio sources.

\begin{table}[hbt!]
\begin{threeparttable}
\caption{Total chi-square values from the best-fits obtained for DES galaxy-auto power spectra in six redshift bins, and their cross-correlation. $b_{g}$ is the best-fit galaxy bias for DES tomographic bins and EMU-PS1 radio sources with model $n(z)$. The $\chi^{2}_{\text{bf}}$ values correspond to our baseline analysis with best-fit model $n(z)$, whereas $\chi^{2}_{\text{TRECS}}$  are using the predicted $n(z)$ from the \texttt{TRECS} simulation. The number of degrees of freedom for $C_{\ell}^{\text{DES}}$ is $\nu=7$ in every redshift bin. For the joint analysis of $C_{\ell}^{\text{EMU}\times \text{DES}}$ and $C_{\ell}^{\text{EMU}}$, $\nu=24$ for \texttt{TRECS}, and $\nu=21$ for model $n(z)$.}
\label{tab:total_chi_square_emu_des}
\begin{tabular}{ccccc}
\toprule
\headrow & & $b_{g}$ & $\chi^{2}_{\text{bf}}$ & $\chi^{2}_{\text{TRECS}}$ \\
\midrule
        & Bin 1 & $1.37^{+0.04}_{-0.04}$ & $9.17$ & - \\
        & Bin 2 & $1.73^{+0.05}_{-0.05}$ & $10.82$ & - \\
$C_{\ell}^{\text{DES}}$ & Bin 3 & $1.74^{+0.05}_{-0.05}$ & $8.08$ & - \\
        & Bin 4 & $1.67^{+0.05}_{-0.05}$ & $8.86$ & - \\
        & Bin 5 & $1.89^{+0.03}_{-0.03}$ & $9.28$ & - \\
        & Bin 6 & $2.08^{+0.03}_{-0.03}$ & $10.19$ & - \\
\midrule
        & Bin 1 & - & $2.22$ & $0.85$ \\
        & Bin 2 & - & $0.37$ & $0.74$ \\
$C_{\ell}^{\text{EMU}\times \text{DES}}$ & Bin 3 & - & $0.46$ & $0.32$ \\
$180\,\mu$Jy & Bin 4 & - & $3.21$ & $3.25$ \\
        & Bin 5 & - & $1.92$ & $1.53$ \\
        & Bin 6 & - & $1.07$ & $1.09$ \\
\midrule
        & Bin 1 & - & $1.51$ & $0.68$ \\
        & Bin 2 & - & $0.44$ & $0.29$ \\
$C_{\ell}^{\text{EMU}\times \text{DES}}$ & Bin 3 & - & $0.32$ & $0.21$ \\
$400\,\mu$Jy & Bin 4 & - & $1.09$ & $1.41$ \\
        & Bin 5 & - & $0.60$ & $0.39$ \\
        & Bin 6 & - & $0.12$ & $0.17$ \\
\bottomrule
\end{tabular}
\end{threeparttable}
\end{table}

\begin{table}[hbt!]
\begin{threeparttable}
\caption{Total chi-square values from the best-fits obtained for joint $C_{\ell}^{\text{EMU}\times \text{DES}}$ spectra, obtained using different models of galaxy bias with our model $n(z)$ The numbers in brackets denote the number of degrees of freedom with each model.}
\label{tab:total_chi_square_emu_des_diff_bias_models}
\begin{tabular}{ccc}
\toprule
\headrow bias model & $180\,\mu$Jy & $400\,\mu$Jy \\
\midrule
    $b(z) = b_{0}/D(z)$ & $35.85$ ($20$) & $11.07$ ($20$) \\
    $b(z) = b_{0}+b_{1}z$ & $14.40$ ($19$) & $4.39$ ($19$) \\
    $b(z) = b_{0}+b_{1}z+b_{2}z^{2}$ & $9.25$ ($18$) & $4.03$ ($18$)  \\
\bottomrule
\end{tabular}
\end{threeparttable}
\end{table}

\section{Conclusion}\label{sec:conclusion}
In this work, we measured the angular power spectra from the cross-correlations between EMU-PS1 radio sources and DES photometric galaxies. By combining these cross-power spectra, we inferred the redshift distribution and galaxy bias of EMU-PS1 radio sources. We validate our findings by fitting the EMU-PS1 auto-power spectrum and its cross-correlation with \textit{Planck} PR3 CMB lensing convergence map.

We applied flux cuts of $180\mu$Jy and $400\,\mu$Jy to the EMU-PS1 source catalogue,  which was generated using the \texttt{PyBDSF} source detection algorithm. We then constructed weights for EMU-PS1 to account for the variations in source counts due to observational noise. Subsequently, we cross-correlated the weighted radio catalogue with the DES MagLim galaxy sample, which was divided into six tomographic bins in the redshift range $0.2 \leq z < 1.05$. The angular power spectra were measured using the \texttt{NaMaster} pseudo-$C_{\ell}$ estimator.

We estimated the galaxy bias in the DES tomographic bins from their auto-power spectrum. For EMU-PS1, we adopted a polynomial galaxy bias model along with a parametric form for its redshift distribution, following \cite{2024A&A...681A.105N}. We also compared the performance of our baseline model $n(z)$ with the widely used simulated redshift distributions from \texttt{TRECS}. The free parameters in our models were constrained using MCMC analysis with \texttt{PyMC}. We prepared mock catalogues of correlated EMU-PS1 and DES surveys to validate our analysis pipeline, as described in Section~\ref{sec:simulations}. Our validation confirmed that our methodology is robust in identifying the peak location of the EMU-PS1 redshift distribution. The key findings of our study are summarized below:
\begin{itemize}
    \item The cross-power spectra between EMU-PS1 and DES bins were detected with $3.8\sigma,\,4.6\sigma,\,3.8\sigma,\,3.5\sigma,\,5.5\sigma$ and $3.9\sigma$ significance, respectively, for $50<\ell<250$.
    \item Our estimates of DES galaxy bias based solely on galaxy clustering measurements are consistent with previous studies except for the last tomographic bin, the reasons for which are discussed in Section \ref{sec:des_gal_bias} (see also \cite{2022PhRvD.105b3520A, 2023PhRvD.107b3530C}).
    \item We found that the \texttt{TRECS} simulation $n(z)$ supplemented with relative contributions from AGN and SFG in each redshift bin adequately reproduce the cross-power spectra in all six DES tomographic bins, as well as the EMU-PS1 auto-power spectrum and cross-power spectrum with CMB convergence. We obtained a total $\chi^{2}$ value of $7.78$ at $180\,\mu$Jy and $3.15$ at $400\,\mu$Jy for \texttt{TRECS}, with $\nu=20$ degrees of freedom.
    \item The EMU-PS1 redshift distribution obtained from our baseline model $n(z)$ (Eq. \ref{eq:model_nz_lotssdr2}) fits all the data equally well, without using any mixture information of radio source populations. Our best-fit model $n(z)$ shows a peak at $z \sim 0.85$, which is higher than the peak of the obtained from \texttt{TRECS} simulations and LOFAR deep fields. However, the peak redshift of our $n(z)$ is consistent with redshift distribution from surveys like COSMOS 3GHz and MIGHTEE.
    \item The galaxy bias evolution at both $180\,\mu$ and $400\,\mu$Jy with our model $n(z)$ is consistent with that obtained from \texttt{TRECS} $n(z)$ and estimates from different works in the literature, without the inclusion of radio source mixture information with \texttt{TRECS} as discussed in Section \ref{sec:results_model}.
\end{itemize}

The goal of this study was to establish cross-correlations with optical surveys as a viable option to validate the redshift distribution and evolution of galaxy bias of radio sources. We have shown that cross-correlation technique, supplemented with simple parametric functions for radio $n(z)$ and galaxy bias, can produce consistent results mitigating the limitations of radio continuum surveys in directly measuring the redshift distribution of sources and accurate classification of radio sources into sub-populations. With additional high-redshift data from optical/near-infrared surveys such as \textit{Euclid} \citep{2025A&A...697A...1E}, the Vera C. Rubin Observatory LSST \citep{2009arXiv0912.0201L, 2019ApJ...873..111I}, and 4MOST \citep{2019Msngr.175....3D}, tighter constraints on the radio source $n(z)$ can be achieved, using both parameteric functions and model independent approach. This, in turn, will enhance the prospects for cosmological studies using the ongoing EMU main survey and the SKA in the future.

\paragraph{Acknowledgments}

The Australian SKA Pathfinder is part of the Australia Telescope National Facility\footnote{\url{https://www.atnf.csiro.au/facilities/askap-radio-telescope/}} which is managed by CSIRO. Operation of ASKAP is funded by the Australian Government with support from the National Collaborative Research Infrastructure Strategy. ASKAP uses the resources of the Pawsey Supercomputing Centre. Establishment of ASKAP, the Murchison Radio-astronomy Observatory and the Pawsey Supercomputing Centre are initiatives of the Australian Government, with support from the Government of Western Australia and the Science and Industry Endowment Fund. We acknowledge the Wajarri Yamatji people as the traditional owners of the Observatory site.

\paragraph{Funding Statement}
BB-K acknowledges support from INAF for the project `Paving the way to radio cosmology in the SKA Observatory era: synergies between SKA pathfinders/precursors and the new generation of optical/near-infrared cosmological surveys' (CUP C54I19001050001).
CLH acknowledges support from the Science and Technology Facilities Council (STFC) through grant ST/Y000951/1 and from the Oxford Hintze Centre for Astrophysical Surveys which is funded through generous support from the Hintze Family Charitable Foundation.
MB is supported by the Polish National Science Center through grants no. 2020/38/E/ST9/00395 and 2020/39/B/ST9/03494.
KT is supported by the STFC grant ST/W000903/1 and by the European Structural and Investment Fund.
SC acknowledges support from the Italian Ministry of University and Research (\textsc{mur}), PRIN 2022 `EXSKALIBUR -- Euclid-Cross-SKA: Likelihood Inference Building for Universe's Research', Grant No.\ 20222BBYB9, CUP D53D2300252 0006, and from the European Union -- Next Generation EU.
JA is supported by MICINN (Spain) grant PID2022-138263NB-I00 (AEI/FEDER, UE) and the Diputaci\'on General de Arag\'on-Fondo Social Europeo (DGA-FSE) Grant No. 2020-E21-17R of the Aragon Government. 

\paragraph{Data Availability Statement}
The DES MagLim data used in this paper is publicly available at \url{https://des.ncsa.illinois.edu/}. The EMU Pilot Survey 1 images and source catalogues can be downloaded from the CSIRO ASKAP Science Data Archive (\url{https://research.csiro.au/casda/}).



\printendnotes

\bibliography{EMU_PS82}

\appendix

\onecolumn
{\section{Full covariance matrix}
The covariance matrix used to fit the cross-power spectra between EMU-PS1 and six tomographic bins from DES Y3 cosmology datasets is given by
\begin{equation}
    {\sf K}_{\ell\ell'} = 
    \begin{bmatrix}
        {\sf K}_{\ell\ell'}^{g_{1}g,g_{1}g} & {\sf K}_{\ell\ell'}^{g_{1}g,g_{2}g} & {\sf K}_{\ell\ell'}^{g_{1}g,g_{3}g} & {\sf K}_{\ell\ell'}^{g_{1}g,g_{4}g} & {\sf K}_{\ell\ell'}^{g_{1}g,g_{5}g} & {\sf K}_{\ell\ell'}^{g_{1}g,g_{6}g}\\
        {\sf K}_{\ell\ell'}^{g_{2}g,g_{1}g} & {\sf K}_{\ell\ell'}^{g_{2}g,g_{2}g} & {\sf K}_{\ell\ell'}^{g_{2}g,g_{3}g} & {\sf K}_{\ell\ell'}^{g_{2}g,g_{4}g} & {\sf K}_{\ell\ell'}^{g_{2}g,g_{5}g} & {\sf K}_{\ell\ell'}^{g_{2}g,g_{6}g}\\
        {\sf K}_{\ell\ell'}^{g_{3}g,g_{1}g} & {\sf K}_{\ell\ell'}^{g_{3}g,g_{2}g} & {\sf K}_{\ell\ell'}^{g_{3}g,g_{3}g} & {\sf K}_{\ell\ell'}^{g_{3}g,g_{4}g} & {\sf K}_{\ell\ell'}^{g_{3}g,g_{5}g} & {\sf K}_{\ell\ell'}^{g_{3}g,g_{6}g}\\
        {\sf K}_{\ell\ell'}^{g_{4}g,g_{1}g} & {\sf K}_{\ell\ell'}^{g_{4}g,g_{2}g} & {\sf K}_{\ell\ell'}^{g_{4}g,g_{3}g} & {\sf K}_{\ell\ell'}^{g_{4}g,g_{4}g} & {\sf K}_{\ell\ell'}^{g_{4}g,g_{5}g} & {\sf K}_{\ell\ell'}^{g_{4}g,g_{6}g}\\
        {\sf K}_{\ell\ell'}^{g_{5}g,g_{1}g} & {\sf K}_{\ell\ell'}^{g_{5}g,g_{2}g} & {\sf K}_{\ell\ell'}^{g_{5}g,g_{3}g} & {\sf K}_{\ell\ell'}^{g_{5}g,g_{4}g} & {\sf K}_{\ell\ell'}^{g_{5}g,g_{5}g} & {\sf K}_{\ell\ell'}^{g_{5}g,g_{6}g}\\
        {\sf K}_{\ell\ell'}^{g_{6}g,g_{1}g} & {\sf K}_{\ell\ell'}^{g_{6}g,g_{6}g} & {\sf K}_{\ell\ell'}^{g_{4}g,g_{3}g} & {\sf K}_{\ell\ell'}^{g_{4}g,g_{4}g} & {\sf K}_{\ell\ell'}^{g_{6}g,g_{5}g} & {\sf K}_{\ell\ell'}^{g_{6}g,g_{6}g}\\
    \end{bmatrix}
    \label{eq:full_cov_matrix}
\end{equation}}
where we used $g$ to denote EMU-PS1 field and $g_{1}-g_{6}$ denote the six DES tomographic bins.

\end{document}